\documentclass[twocolumn,showpacs,preprintnumbers,amsmath,amssymb]{revtex4-1} \usepackage{dcolumn}

\usepackage{graphicx} \usepackage{bm}

\newcommand{\hH}{\hat{H}}

\newcommand{\hP}{\hat{\cal P}^{up}}
\newcommand{\hS}{\hat{S}}
\newcommand{\hU}{\hat{U}}

\begin{document}

\title{Multi-mode behavior of electron Zitterbewegung induced by an electromagnetic wave in graphene}
\date{\today}
\author{Tomasz M. Rusin$^1$}

\email{Tomasz.Rusin@orange.com}

\author{Wlodek Zawadzki$^2$}
\affiliation{$^1$Orange Customer Service sp. z o. o., ul. Twarda 18, 00-105 Warsaw, Poland\\
             $^2$ Institute of Physics, Polish Academy of Sciences, Al. Lotnik\'ow 32/46, 02-688 Warsaw, Poland}

\pacs{72.80.Vp, 42.50.-p, 41.75.Jv, 52.38.-r}

\begin{abstract}
Electrons in monolayer graphene in the presence of an electromagnetic (or
electric) wave are considered theoretically. It is shown that the electron
motion is a nonlinear combination of Zitterbewegung (ZB, trembling motion)
resulting from the periodic potential of graphene lattice and the driving
field of the wave. This complex motion is called ``Multi-mode Zitterbewegung''.
The theory is based on the time-dependent two-band Hamiltonian taking into
account the graphene band structure and interaction with the wave. Our
theoretical treatment includes the rotating wave approximation and high-driving-frequency
approximation for narrow wave packets, as well as
numerical calculations for packets of arbitrary widths. Different
regimes of electron motion are found, depending on relation between the ZB
frequency~$\omega_Z$ and the driving frequency~$\omega_D$ for different strengths
of the electron-wave interaction. Frequencies and intensities of the
resulting oscillation modes are calculated. The nonlinearity of the
problem results in a pronounced multi-mode behavior. Polarization of the
medium is also calculated relating our theoretical results to observable
quantities. The presence of driving wave, resulting in frequencies
directly related to~$\omega_Z$ and increasing the decay time of oscillations,
should facilitate observations of the Zitterbewegung phenomenon.
\end{abstract}

\maketitle

\section{Introduction}
The phenomenon of Zitterbewegung (trembling motion) was devised by Erwin
Schrodinger in 1930~\cite{Schroedinger1930}. Schrodinger remarked that, if one uses the
Dirac equation describing free relativistic electrons in a vacuum, the
velocity operator does not commute with the Dirac Hamiltonian also in
absence of external fields, so that the velocity is not a constant of the
motion. Schrodinger showed that the resulting electron velocity and
trajectory exhibit, in addition to the standard classical components,
also rapidly oscillating components which he called the Zitterbewegung
(ZB). The predicted frequency of ZB oscillations in a vacuum is very high:~$\hbar\omega_Z=2m_0c^2$,
and their amplitude very small:~$\Delta z \simeq \lambda_c=3.86\times 10^{-3}$ \AA . Although the trembling motion has
never been directly observed in a vacuum, the phenomenon has been since
its prediction a source of excitement and subject of studies.
In particular, it was realized that the "reason" for the appearance of ZB
in the Dirac equation is the two-band structure of its energy spectrum.
In 2010 Gerritsma {\it et al.}~\cite{Gerritsma2010} succeeded in simulating the 1+1 Dirac
equation with the resulting Zitterbewegung by means of cold ions
interacting with laser beams.

Since 1970 the trembling motion of charge carriers was also predicted in
superconductors and semiconductors as a consequence two-band energy
spectra in such materials, see the review~\cite{Zawadzki2011}. However, it was not until 2005,
when papers by Zawadzki~\cite{Zawadzki2005KP} and Schliemann {\it et al.}~\cite{Schliemann05} appeared, that the
trembling motion in narrow-gap materials became an intensively
studied subject. In particular, it was clarified that the ``standard''
trembling motion analogous to ZB in a vacuum, is nothing else but an
instantaneous oscillating velocity of an electron moving in a periodic
potential of crystal lattice, see~\cite{Zawadzki2010}. This instantaneous velocity should
be contrasted with an average carrier velocity used in the theory of
transport and optics~\cite{Zawadzki2013}. The trembling motion of charge carriers in
solids has considerably more favorable parameters than that in a vacuum but it
has not been yet observed experimentally, since it is difficult to follow
the motion of a single electron.
After 2005 a real surge of papers on ZB in graphene appeared,
see e.g.~\cite{Katsnelson2006,Cserti2006,Maksimova2008,David2010}.

In order to make the Zitterbewegung more approachable experimentally we
consider in this work the motion of electrons in a periodic potential
subjected in addition to an interaction with an electromagnetic (or
electric) wave. To be specific, we consider electrons in monolayer
graphene because ZB in this material has been studied in some detail~\cite{Rusin2007b},
so that it is easier to follow the influence introduced by the
interaction with the wave. Also, graphene is a material of great current
interest and its electrons are described by a relatively simple~$2 \times 2$
Hamiltonian in which one can neglect the spin~\cite{Wallace1947,Slonczewski1958,McClure1956,Novoselov2005}.
Our work has been inspired by papers treating the Bloch oscillator in the presence of an
electromagnetic wave which has been called
``Super-Bloch oscillator''~\cite{Thommen2002,Kolovsky2009,Haller2010}.
The above problem bears some similarities to our subject since both deal
with materials in which electrons are characterized by an internal
frequency subjected in addition to an interaction with periodic
external perturbation.
The oscillations of average electron velocity
in the presence of a driving electric field are called Multi-mode Zitterbewegung (MZB).

Our subject belongs to a more general domain of problems considering
two-level systems in the presence of periodic perturbations. There exist
quite a few works related to this domain, see e.g.~\cite{AllenBook,SchleichBook}.
However, our treatment has
some specific additional features. First, instead of two levels we deal in
graphene with two two-dimensional bands. Second, when dealing with the
Zitterbewegung one can use two approaches. In the first ZB is
obtained in the Heisenberg picture for time-dependent operators.
This method is usually limited to time-independent Hamiltonians
for which it is possible to solve analytically the Heisenberg equations of motion.
In the second approach one calculates ZB in the Schrodinger picture.
However, the time-dependence of operators does not
represent physical effects and one needs to average results over quantum states.
In this case it is preferable to describe localized electrons in the
form of wave packets since it does not make much sense to consider the
trembling motion of a plane wave having a uniform density in the whole
crystal. We will emphasize differences and similarities of our methods
and results to those of other authors dealing with mathematically related
problems.

Our paper is organized as follows. In Section II we give general
formulation of the problem for an electron in graphene interacting in addition
with a wave. Section III
describes solutions for delta-like packets obtained with the use of
rotating wave and high driving frequency approximations, as well as
complete numerical procedures. Section IV describes motion of Gaussian
wave packets of arbitrary widths and discusses a sub-packet decomposition.
In Section V we consider polarization of the graphene medium which
connects electron motion to observable quantities. In Section VI we
discuss our results. The paper is concluded by a summary. In appendices we
present a short classical description of the main problem, mention its
relation to the Rabi oscillations and compare our results with those
obtained by the approximate Fer-Magnus expansion.

\section{General theory}

We consider monolayer graphene in the presence of a monochromatic electric
wave of the frequency~$\omega_D$ propagating in the~$z$ direction
perpendicular to the graphene sheet. We assume the electric field to be polarized in the~$x$ direction. In the electric dipole
approximation the vector potential of the wave is
\begin{equation} \label{GAt}
A_x(t)=\frac{E_0}{\omega_D}\cos(\omega_D t),
\end{equation}
where~$E_0>0$ is field's intensity, and~$A_y=0$.
The corresponding electric field is~$E_x(t)=-\partial A_x(t)/\partial t=E_0\sin(\omega_Dt)$
and~$E_y=0$. The time-dependent Hamiltonian for charge carriers in monolayer graphene is
\begin{equation} \label{GH}
\hH(t) = u\left(\begin{array}{cc} 0 & \hat{p}_x + eA_x(t) -i \hat{p}_y \\
              \hat{p}_x + eA_x(t) +i \hat{p}_y & 0 \end{array} \right),
\end{equation}
where~$e=|e|$ is the positive electric charge. Because the above Hamiltonian does not depend on~${\bm r}$,
the two-dimensional momentum~${\bm p}=\hbar{\bm k}$ is a good quantum number.
Since we use the electric dipole approximation and do not write explicitly
the magnetic component of the electromagnetic wave interacting with
electron spin, our approach takes into account only the electric component
of the driving wave.
The wave function~$\Psi$ of an electron is a solution of the Schrodinger equation with the time-dependent Hamiltonian
\begin{equation} \label{GHw}
 i\frac{\partial \Psi}{\partial t} = (uk_x \sigma_x + uk_y\sigma_y)\Psi + \omega_e \sigma_x\cos(\omega_Dt)\Psi,
\end{equation}
where~$\omega_e=eE_0u/(\hbar\omega_D)$ and~$\sigma_x$,~$\sigma_y$ are the Pauli matrices.
The first two terms on RHS of Eq.~(\ref{GHw}) correspond to monolayer graphene, while
the last term describes the interaction of charge carriers with the monochromatic wave.

The system described in Eq.~(\ref{GHw}) is characterized by three parameters having frequency dimensions:
driving frequency~$\omega_D$, interband frequency~$\omega_Z=2u|k|$, and strength of the field-matter interaction~$\omega_e$.
The dynamics of the system depends on relative magnitudes of these parameters.
The first step of our calculation is to determine the wave function~$\Psi(t)$ from Eq.~(\ref{GHw}) either exactly,
numerically or with the use of approximate methods. We assume that for~$t=0$ the function~$\Psi(0)$ is a
combination of states with positive and negative energies which is {\it not}
an eigenstate of~$\hH(0)$. Such a state can be prepared by e.g. the ultrashort laser pulse~\cite{Garraway1995}.
Note that in our approach we do not deal with
transitions from the lower to the upper bands but concentrate on the time evolution of~$\Psi(t)$.
We take the initial condition for~$\Psi(0)$ in the form
\begin{equation} \label{GPsi0}
 \Psi({\bm k}, t=0) = f(\bm k)\left(\begin{array}{c} 1 \\ 0 \end{array}\right),
\end{equation}
where
\begin{equation} \label{GGauss}
 f({\bm k})= 2\sqrt{\pi}d e^{-\frac{1}{2}d^2(k_x^2-k_{0x})^2-\frac{1}{2}d^2(k_y-k_{0y})^2}.
\end{equation}
In the real space~$f({\bm r})=\int f({\bm k})e^{-i{\bm k}\cdot{\bm r}} d^2 {\bm k}/(2\pi)^2$ describes the Gaussian wave packet
of a width~$d$ and the initial quasi-momentum~$(\hbar k_{0x}, \hbar k_{0y})$. We also consider a delta-like packet obtained
from the Gaussian packet as the limit
\begin{equation} \label{GDelta}
 \lim_{d\to \infty} |f({\bm k})|^2 \rightarrow \delta(k_x-k_{0x})\delta(k_y-k_{0y}).
\end{equation}
Note that we take the limit for~$|f({\bm k})|^2$ instead for~$f({\bm k})$ in order to avoid squares of the Dirac-delta
function in the matrix elements of velocity operators, see below.

In the following we analyze oscillations of the packet motion and concentrate on the average velocity rather than position.
For time independent Hamiltonian the velocity operator is:~$\hat{v}_i = \partial \hH/\partial \hat{p}_i$,
and the average velocity is given by the
matrix element of~$\hat{v}_i$ averaged on the wave packet of Eq.~(\ref{GGauss}). This procedure may not be applied to the
time-dependent Hamiltonian of Eq.~(\ref{GH}) and the average packet velocity must be calculated differently.
Let us consider an average electron position
\begin{equation} \langle x(t) \rangle = \langle \Psi(t)|x|\Psi(t)\rangle, \end{equation}
where~$\Psi(t)$ is a solution of Eq.~(\ref{GHw}). Upon using twice Eq.~(\ref{GHw}) one finds
\begin{eqnarray} \label{Gdxdt}
 \frac{d\langle x(t) \rangle}{dt} &=& \left \langle \frac{d\Psi(t)}{dt}\left|x\right|\Psi(t)\right\rangle
   + \left\langle \Psi(t)|x|\frac{d\Psi(t)}{dt} \right\rangle \nonumber \\
 &=& \frac{i}{\hbar} \langle \Psi(t)|[\hH, x]|\Psi(t) \rangle.
\end{eqnarray}
The above equation resembles the usual time dependence of operators in the Heisenberg picture.
However, it is valid {\it only} for the matrix elements between the solutions~$\Psi(t)$ of the
Schrodinger equation~(\ref{GHw}).
Calculating the commutator in Eq.~(\ref{Gdxdt}) we obtain
\begin{eqnarray}
\label{Gvx}
 \langle v_x(t) \rangle &=& u \langle \Psi(t)|\sigma_x|\Psi(t)\rangle, \\
 \label{Gvy}
 \langle v_y(t) \rangle &=& u \langle \Psi(t)|\sigma_y|\Psi(t)\rangle.
\end{eqnarray}
In the next sections we analyze the average velocity as a function of~$\omega_Z$,~$\omega_D$ and~$\omega_e$.
First, we find general properties of the electron motion. Calculating the
time derivatives of~$\langle v_x(t)\rangle$ and~$\langle v_y(t)\rangle$ with the method used in Eq.~(\ref{Gdxdt}), we obtain
\begin{eqnarray}
\label{Gax}
\frac{d\langle v_x \rangle}{dt} &=& -2k_y u^2 \langle\sigma_z\rangle, \\
 \label{Gay}
 \frac{d\langle v_y\rangle }{dt} &=& 2u[uk_x+\omega_e\cos(\omega_Dt)]\langle\sigma_z\rangle.
\end{eqnarray}
One sees from Eq.~(\ref{Gax}) that for~$k_y=0$ there is~$d\langle v_x(t) \rangle/dt=0$, which means that the
electron moves with a constant velocity. For~$\langle v_x(0)\rangle=0$ there is no packet motion in the~$x$ direction,
which agrees with ZB results for the field-free case~\cite{Rusin2007b}. Taking~$k_x=0$ in Eq.~(\ref{Gay})
we obtain a nonzero~$\langle v_y(t) \rangle$.
Finally, we calculate the third time-derivative of~$\langle v_x \rangle$,
which is proportional to the second time-derivative of~$\langle \sigma_z \rangle$, see Eq.~(\ref{Gax}). One has
\begin{eqnarray} \label{Gztt}
\frac{d^2\langle \sigma_z \rangle}{dt^2} &=& 4\left\{[uk_x+\omega_e\cos(\omega_Dt)]^2 + u^2k_y^2 \right \}\langle \sigma_z \rangle \nonumber \\
               &&  - i\omega_D\omega_e\sin(\omega_Dt)\langle v_y\rangle.
\end{eqnarray}
For~$\omega_e=0$ there are no time-dependent terms on RHS of Eq.~(\ref{Gztt}) and~$\langle \sigma_z \rangle$
oscillates with one frequency~$\omega_Z=2u\sqrt{k_x^2+k_y^2}$. Substituting~$\langle \sigma_z \rangle$
to Eqs.~(\ref{Gax}) and~(\ref{Gay}) we obtain ZB in the field-free case.
For a non-zero field we take for simplicity~$k_x=0$ and a low driving
frequency:~$\omega_D \ll \omega_e, \omega_Z$. In this approximation we may neglect the last term in Eq.~(\ref{Gztt}) and, upon
using the identity~$\cos^2(x)=(1+\cos(2x))/2$, obtain
\begin{equation}\label{GzMathieu}
\frac{d^2\langle \sigma_z \rangle}{d\tau^2}\simeq \left\{2(E_0e)^2\cos(2\tau) + (2uk_y\omega_D)^2 + 2(E_0e)^2 \right \}\langle \sigma_z \rangle,
\end{equation}
where~$\tau=\omega_Dt$. For~$E_0=0$, Eq.~(\ref{GzMathieu}) reduces to the harmonic oscillator equation.
For~$E_0>0$ the time dependence of~$\langle \sigma_z(\tau) \rangle$ is given by the Mathieu equation,
whose solutions are periodic functions oscillating with many frequencies.
In the general case they can not be expressed by elementary functions.
The analysis presented above indicates that the presence of a monochromatic
electric wave introduces more than one oscillating component.
Below we analyze this aspect in more detail.

\section{Delta-like packet}

We analyze first the motion of a delta-like packet centered at~${\bm k}_0=(0,k_{0y})$ using Eqs.~(\ref{GPsi0}) and~(\ref{GDelta}).
For such a packet our situation resembles a two-level system driven by a periodic force.
We begin the analysis with two approximations: the rotating wave approximation and
the high-driving-frequency approximation, which were frequently applied to problems described by the two-level system.

\subsection{Rotating wave approximation (RWA)}

\begin{figure}
\includegraphics[width=8.0cm,height=8.0cm]{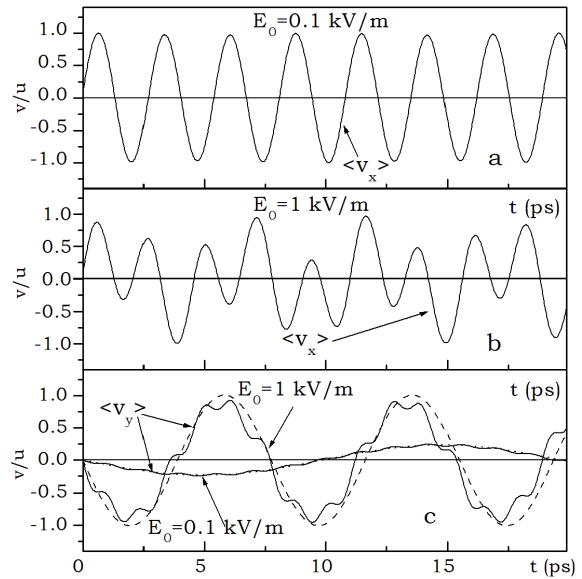}
\caption{Average velocity of a delta-like packet in~$x$ and~$y$ directions versus time calculated
         numerically (solid lines) and within RWA (dashed lines). For~$\langle v_x(t) \rangle$ the exact and RWA results are indistinguishable.
         Model parameters:~$\omega_D=2\times 10^{12}$~s$^{-1}$ and~$\omega_Z=2.31\times 10^{12}$~s$^{-1}$.} \label{Fig1}
\end{figure}

\begin{figure}
\includegraphics[width=8.0cm,height=8.0cm]{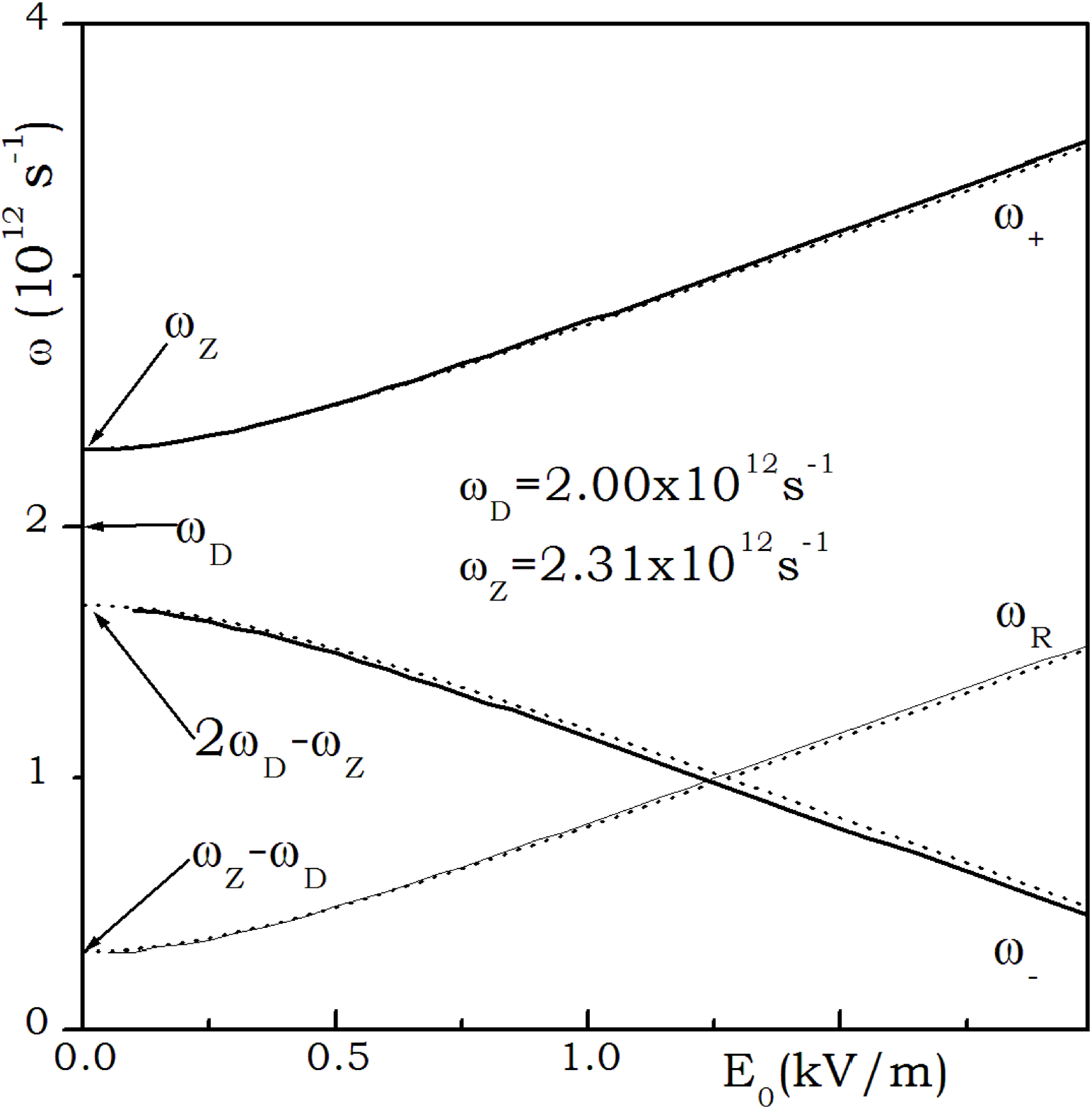}
\caption{Frequencies of motion components for a delta-like wave packet
         versus electric field of EM wave. Upper line: ZB-like component of~$\langle v_x \rangle$;
         middle line: satellite component of~$\langle v_x \rangle$; lower line:~$\langle v_y \rangle$ mode.
         Solid lines: exact results, dotted lines: approximations given by RWA.} \label{Fig2}
\end{figure}

\begin{figure}
\includegraphics[width=8.0cm,height=8.0cm]{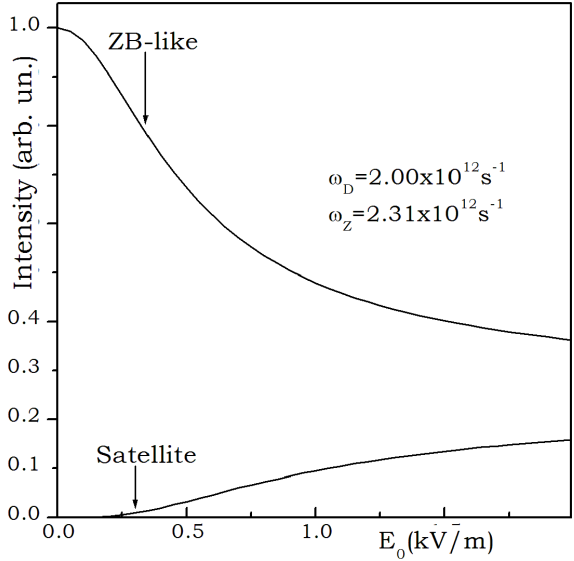}
\caption{Intensities of ZB-like and satellite modes of~$\langle v_x(t)\rangle$ calculated within RWA for a delta-like packet.} \label{Fig3}
\end{figure}

The rotating wave approximation (RWA) allows one to find approximate solutions of Eq.~(\ref{GHw})
when~$\omega_D \simeq \omega_Z$ and the electric field~$E_0$ is weak.
We calculate average packet velocities in a few steps. First we note that, for a delta-like packet centered at~${\bm k}_0=(0,k_{0y})$,
one may set~$k_x=0$, which removes the~$k_x$ dependence of the Hamiltonian in Eq.~(\ref{GHw}).
Next we transform Eq.~(\ref{GHw}) with the unitary transformation:~$\hS_x=(1-i\sigma_x)/\sqrt{2}$. This gives
\begin{equation} \label{RWA_Schpsi}
 i\frac{d\psi}{dt} = \left\{ \omega_e\cos(\omega_Dt)\sigma_x + uk_y\sigma_z\right\}\psi,
\end{equation}
where~$\psi=\hS_x\Psi$. The initial condition is~$\psi(0)=(1,-i)^T/\sqrt{2}$.
We assume solutions of Eq.~(\ref{RWA_Schpsi}) to be of the form
\begin{equation} \label{RWA_psia}
 \psi(t) = \left( \begin{array}{c} \psi_1(t) \\ \psi_2(t)\end{array} \right) =
  \left( \begin{array}{c} a_1(t)e^{-iuk_yt} \\ a_2(t)e^{+iuk_yt}\end{array} \right),
\end{equation}
where~$a_1(t)$ and~$a_2(t)$ are unknown functions. On substituting~$\psi(t)$ into Eq.~(\ref{RWA_Schpsi}) and
neglecting terms oscillating with the frequency~$\omega_Z + \omega_D$ one obtains
\begin{eqnarray}
 \label{RWA_RWA1}
\frac{da_1}{dt} &\simeq& -i\frac{\omega_e}{2} e^{+i\Delta t}a_2, \\
 \label{RWA_RWA2}
\frac{da_2}{dt} &\simeq& -i\frac{\omega_e}{2} e^{-i\Delta t}a_1,
\end{eqnarray}
where~$\Delta=\omega_Z - \omega_D$ with~$\omega_Z=2uk_y$ being the interband ZB frequency.
After some algebra we find
\begin{eqnarray} \label{RWA_psi_w}
\psi_1(t) &=& C_+ e^{-i\omega_D t/2 + i\omega_R t/2} + C_- e^{-i\omega_D t/2 - i\omega_R t/2}, \ \ \ \ \ \\
\psi_2(t) &=& D_+ e^{+i\omega_D t/2 + i\omega_R t/2} + D_- e^{+i\omega_D t/2 - i\omega_R t/2}, \ \ \ \ \
\end{eqnarray}
where~$C_{\pm}$ and~$D_{\pm}$ are constants determined from the initial condition, and
\begin{equation} \label{RWA_OmegaR}
 \omega_R = \sqrt{\left(\omega_Z-\omega_D \right)^2+ \left(\frac{eE_0u}{\hbar\omega_D}\right)^2}.
\end{equation}
The obtained frequency~~(\ref{RWA_OmegaR}) corresponds to the generalized Rabi frequency appearing in the
oscillating occupation of periodically driven two-level systems~\cite{Rabi1937,BoydBook}.
Having~$\psi(t)=\hS_x\Psi(t)$ we calculate the average velocities, see Eqs.~(\ref{Gvx}) and~(\ref{Gvy})
\begin{eqnarray} \label{RWA_vx_w}
\langle v_x(t) \rangle &=& \frac{u}{2} \left(1+\frac{\Delta}{\omega_R}\right)\sin(\omega_Dt+\omega_Rt ) \nonumber \\
                    &+& \frac{u}{2} \left(1-\frac{\Delta}{\omega_R}\right)\sin(\omega_Dt-\omega_Rt ), \\
                 \label{RWA_vy_w}
\langle v_y(t) \rangle &=& -u\frac{\omega_e}{\omega_R}\sin(\omega_Rt).
\end{eqnarray}
Equation~(\ref{RWA_vx_w}) describes electron motion oscillating with two frequencies:
\begin{equation} \label{RWA_Omega_pm}
 \omega_{\pm} = \omega_D \pm \omega_R,
\end{equation}
Let us consider the case of small electric fields and~$\omega_D < \omega_Z$.
From Eq.~(\ref{RWA_OmegaR}) there is~$\omega_R \simeq \omega_Z-\omega_D$, so that
the frequency in the first line of Eq.~(\ref{RWA_vx_w}) is~$\omega_D+\omega_R\simeq \omega_Z$,
i.e. it corresponds to the Zitterbewegung frequency
in the field-free case. The second frequency (satellite, SAT) equals then to~$2\omega_D-\omega_Z$.
Since~$\Delta=\omega_Z-\omega_D>0$, the amplitude of the satellite approaches zero, while the amplitude of
ZB part approaches unity. For~$\omega_D > \omega_Z$ the roles of the first and second contributions to the velocity are
reversed. The~$y$ component of motion oscillates with the generalized Rabi frequency~$\omega_R$
and its amplitude is proportional to the field intensity.
In general, the~$\langle v_y(t) \rangle$ component is not related to the ZB oscillations.
However, the Rabi frequency includes the ZB frequency~$\omega_Z$, so
at low fields~$E_0$ one can measure~$\omega_Z$ from the relation~$\omega_R\simeq|\omega_Z-\omega_D|$.

Results for~$\langle v_x(t)\rangle$ calculated for two electric fields are plotted in Fig.~1a and Fig.~1b,
while in Fig.~1c we show results for~$\langle v_y(t)\rangle$.
The solid lines show the results obtained numerically
by solving Eq.~(\ref{GHw}) with the use of fifth-order Runge-Kutta method,
while the dashed lines are obtained within RWA from Eq.~(\ref{RWA_vy_w}).
Here the parameters are within the range of validity of RWA and the average velocity~$\langle v_x(t) \rangle$,
as calculated numerically, is undistinguishable from the motion obtained with the use of Eq.~(\ref{RWA_vx_w}).
However, for larger fields there exist deviations from the RWA results for~$\langle v_y (t)\rangle$,
because for~$\langle v_y (t)\rangle$ RWA predicts one frequency of oscillations,
see Eq.~(\ref{RWA_vy_w}), while the numerical calculations predict a small additional modulation.

The average packet velocity
oscillates from~$-u$ to~$u$ and the motion does not disappear in time.
The amplitude of~$\langle v_x(t)\rangle$ does not depend on electric field.
The frequencies of both ZB and SAT oscillations
depend on the electric field and the frequency of the ZB mode differs from~$\omega_Z=2uk_{0y}$.
Patterns presented in Fig.~1a, 1b, and 1c are periodic in time.
Using Fig.~1 we may define the Multi-mode Zitterbewegung as
an {\it appearance of additional oscillation modes} in electron motion in the presence of an electromagnetic wave.

It is seen in Fig.~1c that the packet motion in the~$y$ direction differs significantly from that in the~$x$ direction.
First, the amplitude of~$\langle v_y(t)\rangle$ depends on
the electric field: it vanishes for low fields and grows to~$\pm u$ for large fields.
Second,~$\langle v_y(t)\rangle$ oscillates with
one frequency for both low and high fields.
Third, the amplitude of~$\langle v_y(t)\rangle$ and the oscillation frequency
strongly depends on the electric field. Finally, the oscillation frequency
of~$\langle v_y(t)\rangle$ is not related to~$\omega_Z$.

Figure~2 shows frequencies of~$\langle v_x(t)\rangle$ motion as functions of electric field~$E_0$.
The upper line describes the ZB-related frequency,
while the middle line shows the satellite frequency.
The lowest line presents the results for~$\langle v_y(t)\rangle$. Numerical values are indicated
by the dotted lines, the solid lines are obtained from the RWA formulas presented above.
There is a very good agreement between the numerical results and those predicted by RWA.

Calculating the Fourier transform of~$\langle v_x(t)\rangle$ we obtain the frequency
spectrum~$\langle v_x(\omega)\rangle$ of the motion. The intensities~$I_x(\omega)$ of
motion components are proportional to~$|\langle v_x(\omega)\rangle|^2$.
In Fig.~3 we show intensities of the ZB-like and satellite components determined by RWA.
For~$E_0=0$ the intensity of satellite vanishes and the packet oscillates with one frequency corresponding to
the ZB frequency in the field-free case. For small fields, the ZB-like
component dominates over the satellite component.
By increasing the field one observes a gradual decrease of the ZB intensity and a slow growth of the
satellite intensity. For still larger fields the intensities of both components are comparable.

\subsection{High driving frequency (HDF)}

To describe the electron motion for high driving frequency~$\omega_D$ we consider again a delta-like wave
packet centered around~${\bm k}_0=(0,k_{0y})$. By taking~$k_x=0$ and transforming Eq.~(\ref{GHw})
with the use of unitary operator~$\hS_y=(1+i\sigma_y)/\sqrt{2}$ one obtains
\begin{equation} \label{HDF_Schphi}
i \frac{d\phi}{dt} = \left\{ \omega_e\cos(\omega_Dt)\sigma_z + uk_y\sigma_y\right\}\phi,
\end{equation}
where~$\phi=\hS_y\Psi$. The initial condition for~$\phi$ is:~$\phi(0)=(1,-1)^T/\sqrt{2}$.
We assume solutions of Eq.~(\ref{HDF_Schphi}) to be in the form
\begin{equation} \label{HDF_phib}
\phi(t) = \left( \begin{array}{c} b_1(t)e^{-i\kappa \sin(\omega_Dt)} \\ b_2(t)e^{+i\kappa \sin(\omega_Dt)}\end{array} \right),
\end{equation}
where~$\kappa=\omega_e/\omega_D$ and~$b_1(t)$,~$b_2(t)$ are unknown functions.
On substituting~$\phi(t)$ into Eq.~(\ref{HDF_Schphi}) one has
\begin{eqnarray}
\label{HDF_Schb1}
 i\frac{db_1(t)}{dt} &=& -iuk_y e^{+2i\kappa \sin(\omega_Dt)}b_2(t), \\
\label{HDF_Schb2}
 i\frac{db_2(t)}{dt} &=& +iuk_y e^{-2i\kappa \sin(\omega_Dt)}b_1(t).
\end{eqnarray}
The above equations are still exact. For~$\kappa \ll 1/2$
we approximate~$e^{\pm 2i\kappa \sin(\omega_Dt)} \simeq 1$ and the above equations reduce to
\begin{eqnarray}
\label{HDF_Schb1apr}
 i\frac{db_1(t)}{dt} &\simeq& -iuk_y b_2(t), \\
\label{HDF_Schb2apr}
 i\frac{db_2(t)}{dt} &\simeq& +iuk_y b_1(t),
\end{eqnarray}
which can be easily solved. Using Eq.~(\ref{HDF_Schphi}) one obtains
\begin{eqnarray} \label{HDF_phi_w}
\phi_1(t) &=&  \frac{1-i}{2}e^{+i\omega_k t-i\Lambda}+\frac{1+i}{2}e^{-i\omega_k t-i\Lambda}, \\
\phi_2(t) &=& -\frac{1+i}{2}e^{+i\omega_k t+i\Lambda}-\frac{1-i}{2}e^{-i\omega_k t+i\Lambda},
\end{eqnarray}
where~$\Lambda(t) = \kappa \sin(\omega_Dt)$. Using Eqs.~(\ref{Gvx}) and~(\ref{Gvy}) we find
components of the average velocity
\begin{eqnarray} \label{HDF_vx}
\langle v_x(t)\rangle &=& u\sin(\omega_Zt), \\
 \label{HDF_vy}
\langle v_y(t)\rangle &=& u\sin(2\Lambda)\cos(\omega_Zt) \nonumber \\
                      &\simeq&u\frac{\omega_e}{\omega_D}[\sin(\omega_Dt-\omega_Zt) + \sin(\omega_Dt+\omega_Zt)].
\end{eqnarray}
For high driving frequency the average velocity~$\langle v_x(t)\rangle$ oscillates with one frequency~$\omega_Z$
as in field-free case.
One can interpret this result by saying that the electron cannot follow such a high
frequency, so the effect of the wave averages to zero.
The second motion component oscillates with two frequencies~$\omega=\omega_D\pm \omega_Z$.
Validity of the above approximation requires a small value of the parameter
\begin{equation} \label{HDF_kappa}
\kappa = \frac{\omega_e}{\omega_D} = \frac{eE_0u}{\hbar\omega_D^2} \ll \frac{1}{2}.
\end{equation}
Therefore, for fixed~$E_0$, the parameter~$\kappa$ decreases {\it quadratically} with~$\omega_D$.

\subsection{Numerical results}

\begin{figure}
\includegraphics[width=8.0cm,height=8.0cm]{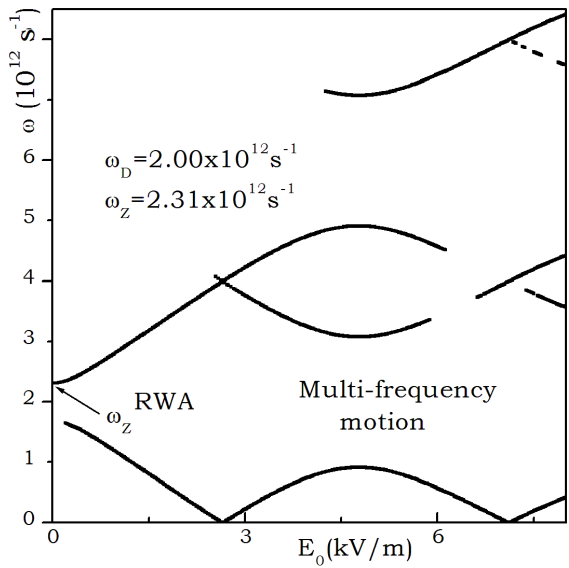}
\caption{Main frequencies of MZB motion calculated for a delta-like wave packet vs.
         electric field of the wave in~$\omega_D \simeq \omega_Z$ regime. } \label{Fig4}
\end{figure}

From Fig.~3 we see that for~$E_0\ge$ 1 kV/m the amplitudes of both motion components are comparable.
Thus a strong driving wave changes a single-frequency motion into a two-frequency motion.
It is expected that still higher electric fields lead to appearance of even more components.
To verify this expectation we calculate numerically the average packet velocity
for large electric fields. Next we carry out the
Fourier transform of~$\langle v_x(t)\rangle$ and identify main frequencies in the motion.

For given~$\bm{k}=(0,k_{0y})$, the numerical calculations of the wave function~$\Psi(t)$ in Eq.~(\ref{GHw})
are performed using the fifth-order Runge-Kutta method. Having determined~$\Psi(t)$ we calculate
the average velocity with the use of Eq.~(\ref{Gvx}) as a function of electric field~$E_0$.
The results are shown in Fig.~4. For each value of~$E_0$ we plot components of~$\langle v_x(t)\rangle$
for which the absolute value of its Fourier transform exceeds the threshold defined below.
Assuming that the Fourier spectrum~$\langle v_x(\omega)\rangle$ is normalized as:~$\int |\langle v_x(\omega)\rangle| d\omega=1$
we take the threshold at the level:~$|\langle v_x(\omega)\rangle| >0.05$.
For low electric fields the packet motion has two frequencies and it is well described by RWA.
The field intensity corresponding to
apparent zero frequency (just below~$E_0=$ 3kV/m) determines the upper range of validity of RWA.
For still larger fields a nonlinear wave mixing appears and the number of motion
frequencies increases to three, four, etc. Note that the additional frequencies appear {\it gradually},
i.e. their amplitudes grow from zero. They are plotted in Fig.~4 when their intensities exceed the threshold mentioned above.
The same can be said about the bifurcation point close to~$E_0\simeq$ 3 kV/m:
the lower branch appears gradually and the upper branch gradually disappears.

\begin{figure}
\includegraphics[width=8.0cm,height=8.0cm]{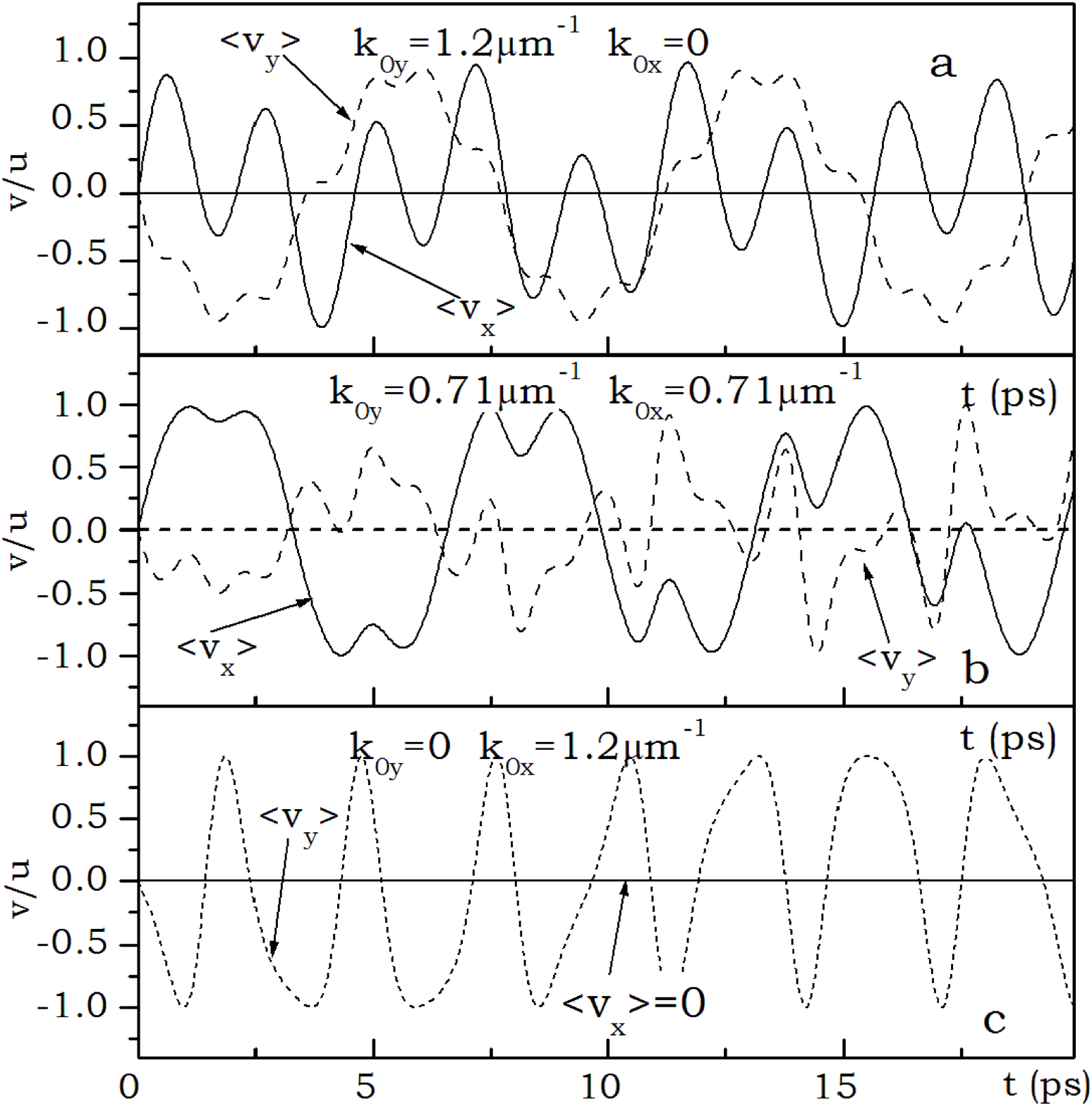}
\caption{Average velocities~$\langle v_x(t) \rangle$ and~$\langle v_y(t) \rangle$ of a delta-like packet with different
         values of~$k_{0y}$ and~$k_{0x}$.
         Electric field is~$E_0=1$~kV/m and~$\omega_D=2\times 10^{12}$~s$^{-1}$.
         For~$k_{0y}=0$ there is no motion in~$x$ direction.} \label{Fig5}
\end{figure}

Until now we concentrated on the delta-like wave packet with a nonzero wave vector in the~$y$ direction.
Now we analyze the Multi-mode ZB for the delta-like
packet with an arbitrary direction of the initial wave vector, i.e.~${\bm k}_{0}=(k_{0x},k_{0y})$.
In the calculations we keep the same length of the wave vector, but change
its direction. Equation~(\ref{GHw}) is solved numerically for three values of~${\bm k}_0$,
the results are plotted in Fig.~5. Comparing maxima or minima of~$\langle v_x(t)\rangle$ in Fig.~5a and Fig.~5b
we note that the main oscillation frequencies remain almost unchanged. Finally, when~$k_{0y}=0$, there is no electron motion in
the~$x$ direction but there is still motion in the~$y$ direction, see Fig.~5c.
This result is in agreement our earlier predictions, see Eqs.~(\ref{Gax}) and~(\ref{Gay}).

\begin{figure}
\includegraphics[width=8.0cm,height=8.0cm]{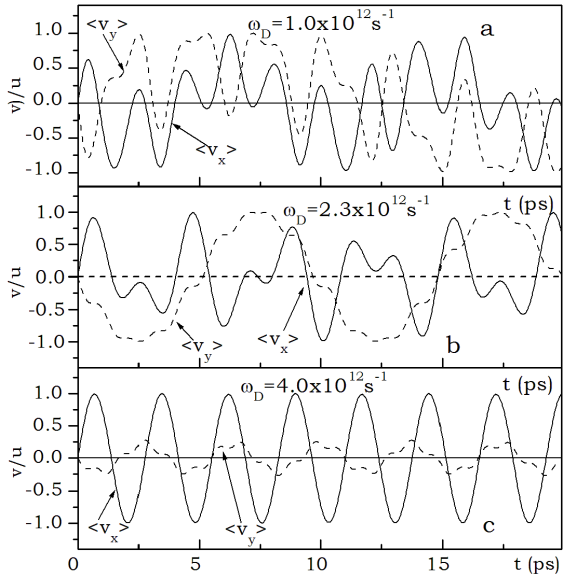}
\caption{Average velocities~$\langle v_x(t) \rangle$ and~$\langle v_y(t) \rangle$ calculated for a
        delta-like packet for three values of~$\omega_D$. Electric field is~$E_0=1$~kV/m.} \label{Fig6}
\end{figure}

In Fig.~6 we show numerical results for another arrangement, when one keeps the electric field constant but
changes the driving frequency. When~$\omega_D$ is much smaller than~$\omega_Z$, see Fig.~6a,
the motion consists of many modes. When both frequencies are equal, see Fig.~6b,
the motion has two well defined frequencies.
For~$\omega_D$ larger than~$\omega_Z$, see Fig.~6c, the electron oscillates with {\it one} frequency equal to~$\omega_Z$,
the same as in the field-free case. This situation corresponds to the HDF case analyzed in the previous
subsection and the numerical calculations confirm the validity of the HDF approximation.

\begin{figure}
\includegraphics[width=8.0cm,height=8.0cm]{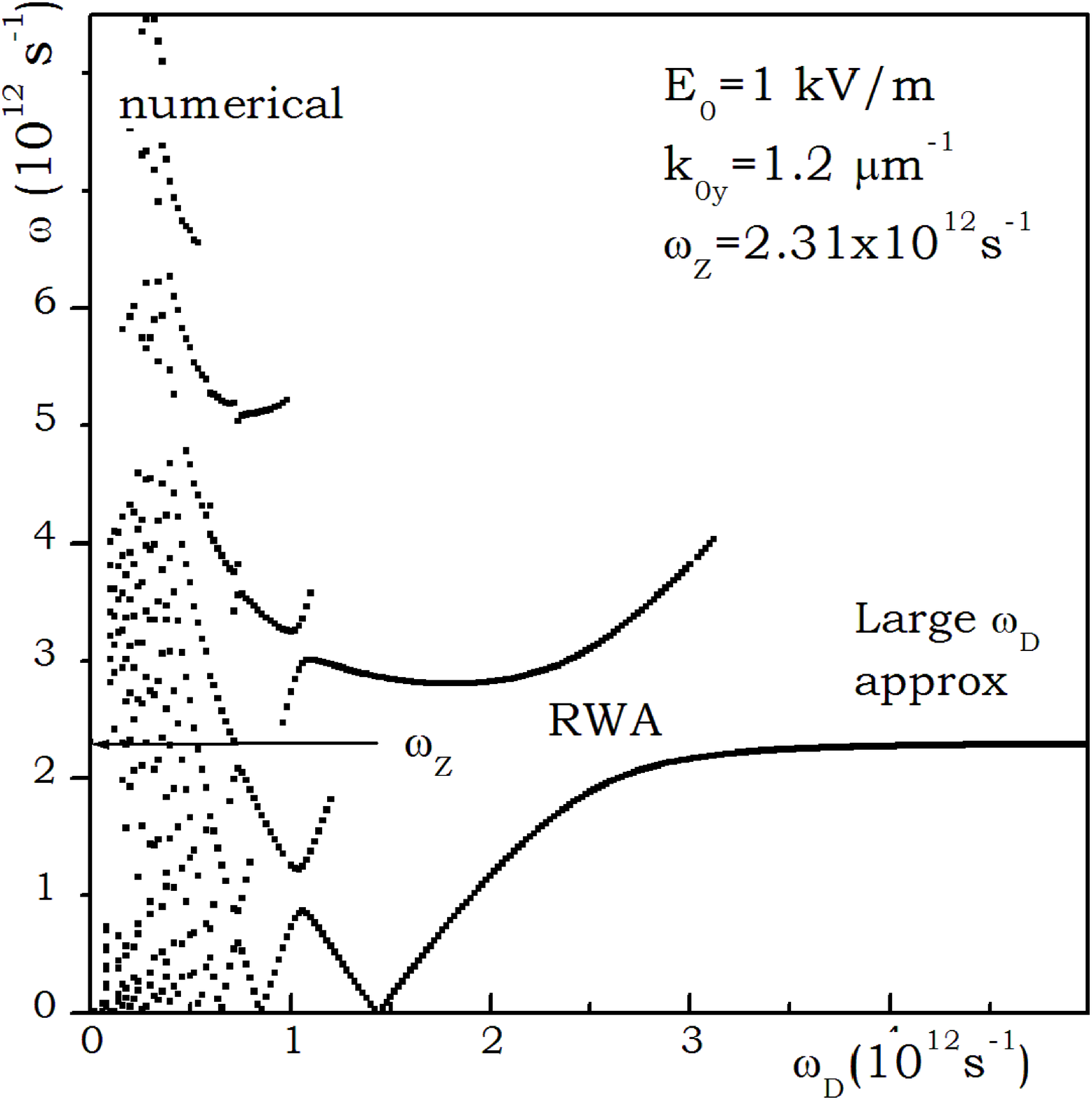}
\caption{Main frequencies of MZB motion calculated for a delta-like wave packet vs. driving frequency~$\omega_D$.
         Three regimes are observed:~$\omega_D \gg \omega_Z$ (one mode),~$\omega_D \propto \omega_Z$ (two modes)
         and~$\omega_D \ll \omega_Z$ (many modes).} \label{Fig7}
\end{figure}

In Fig.~7 we show the spectrum of Multi-mode ZB motion in the~$x$ direction as a function of~$\omega_D$ for fixed~$E_0$.
This figure summarizes the basic properties of the Multi-mode ZB. For large~$\omega_D$ there is only one frequency,
the same as in the field-free case. For~$\omega_D \simeq \omega_Z$
there are two modes oscillating with modified ZB and satellite frequencies.
This is the region of validity of RWA. The latter works correctly until one of the two frequencies goes to zero.
Then the nonlinear wave mixing appears and the motion consists of modes oscillating
with many frequencies. In this region it is possible to identify a limited range of~$\omega_D$ giving two-band descriptions
of the motion, as e.g. for the two lower branches near~$\omega_D=1.1\times 10^{12}$~s$^{-1}$.
The complicated structure of MZB in this region
makes it difficult to identify the~$\omega_Z$ frequency. The conclusion from Figs.~4 and~7 is
that the most promising regions for experimental
observation of ZB are: 1) small electric fields for ~$\omega_D \simeq \omega_Z$
when RWA is valid, 2) large~$\omega_D$ when one ZB frequency occurs.

\section{Motion of Gaussian packet}

The above analysis uses a narrow delta-like packet in~${\bm k}$ space including only one wave vector~${\bm k}={\bm k}_0$.
The obtained results allow one to identify main features of MZB and to establish connection between numerical
results and RWA or large~$\omega_D$ approximations. However, weakness of the delta-like packets
is that the latter are completely delocalized in space and their oscillations have a limited sense. For this reason one should
consider packets of finite widths, which are well localized both in~${\bm k}$ and real spaces. A finite size of the packet
allows one to calculate the average position and its average velocity interpreted as the group velocity.
As shown previously~\cite{Rusin2007b}, a finite packet does not alter main features of ZB: oscillation frequency
remains nearly constant for various packet widths and for wide packets (in real space) the amplitude
of first oscillations is almost the same as that for delta-like packets. We expect the same similarities to occur for MZB.
As pointed out by Lock~\cite{Lock1979}, the wave packet oscillations disappear in time as a result of the Riemann-Lebesgue theorem.
In consequence, the ZB motion has a transient character. The decay of motion is caused by the presence of
wave packet and strongly depends on packet's width~$d$~\cite{Rusin2007b}.
The presence of electric wave introduces two additional parameters to the system: field intensity~$E_0$ and
driving frequency~$\omega_D$ which alter, among other features, the decay time of the packet.
As we show in next subsections, the decay times of MZB oscillations are longer than those for ZB alone.

\subsection{Average velocity}

\begin{figure}
\includegraphics[width=8.0cm,height=8.0cm]{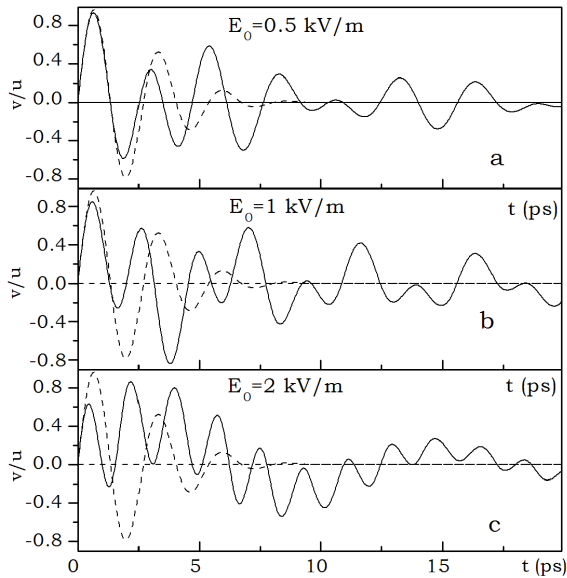}
\caption{Average velocity~$\langle v_x(t) \rangle$ of Gaussian packet for three electric fields.
         Driving frequency is~$\omega_D=2\times 10^{12}$~s$^{-1}$ and~$\omega_Z=2.31\times 10^{12}$~s$^{-1}$.
         Packet parameters:~$d=4\mu$m;~$k_{0x}=0$;~$k_{0y}=1.2 \mu$m$^{-1}$.
         Dashed lines:~$\langle v_x(t) \rangle$ in absence of fields.} \label{Fig8}
\end{figure}

\begin{figure}
\includegraphics[width=8.0cm,height=8.0cm]{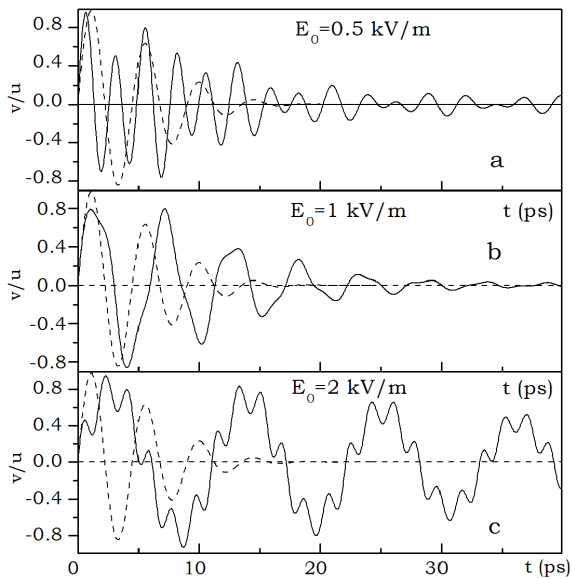}
\caption{The same as in Fig.~8 but for a wider packet.
         Driving frequency is~$\omega_D=2.0\times 10^{12}$~s$^{-1}$ and~$\omega_Z=1.39\times 10^{12}$~s$^{-1}$.
         Packet parameters:~$d=8\mu$m;~$k_{0x}=0$,~$k_{0y}=0.71\mu$m$^{-1}$.
         Dashed lines:~$\langle v_x(t) \rangle$ in absence of fields.} \label{Fig9}
\end{figure}

Average packet velocities are calculated numerically using Eqs.~(\ref{GHw}),~(\ref{Gvx}) and~(\ref{Gvy}).
First, we select the packet width~$d$ and find a rectangle in the~${\bm k}$
space in which the Gaussian wave function in Eq.~(\ref{GGauss}) does
not vanish. In this rectangle we make a two-dimensional grid~$N_x \times N_y$ of different~${\bm k}=(k_x,k_y)$ values.
Next, for each point of the grid we solve numerically Eq.~(\ref{GHw}) using the fifth-order Runge-Kutta method.
Finally, having~$\Psi_{\bm k}(t)$ tabulated for all points in the grid and at all instants of time~$t$,
we calculate numerically, for each instant~$t$, the double integrals over~$d^2{\bm k}$ involved in
average velocities in Eqs.~(\ref{Gvx}) and~(\ref{Gvy}). The integration is performed over all points of the grid in a standard way.
Strong localization of the Gaussian packet in~${\bm k}$ space and proper normalization of~$\Psi(t)$
for every~$t$ ensures the convergence of the integrals.
The results of all calculations presented in the next sections are obtained
for~$N_x=N_y=141$, i.e. for a given packet width Eq.~(\ref{GHw}) is solved 19881 times.
We checked the accuracy of our calculations by increasing the grid up to~$N_x=N_y=201$. Within the presented range of packet parameters
the results are unchanged.

In Figs.~8 and~9 we show the MZB motion for two sets of packet parameters and four values of external driving field.
The wave packet at~$t=0$ is given in Eq.~(\ref{GGauss}).
Dashed lines indicate the field-free ZB motion, solid lines correspond to the motion in the presence of
field (MZB). The parameters in Fig.~8 are in the region of validity of RWA.
For a small field~$E_0$, see Fig.~8a, the MZB oscillations follow pure ZB in terms of the oscillation frequency and amplitude, but
they have much longer decay time. For larger fields, the MZB motion differs qualitatively and quantitatively from pure ZB oscillations.
First, the MZB frequency is larger than for ZB. Second, MZB follows its delta-packet description (see Eq.~(\ref{RWA_vx_w}))
and oscillates with {\it two} frequencies. Third, the decay time of MZB is much longer than the corresponding time of ZB,
difference between the two is more than one order of magnitude.
An intuitive explanation of the last difference is that the driving force, which persists in time, mixes with the transient
ZB oscillation leading to a prolongation of MZB.
A longer decay time of MZB oscillations should make a possible experimental detection of MZB easier.

The results shown in Fig.~9 are obtained for the driving frequency significantly larger than that of ZB. We
observe a gradual transition from a two-frequency motion for low fields, see Fig.~9a,
to a multi-frequency motion for large
fields shown in Fig.~9c. The change in packet motion is similar to that shown in Fig.~4.
Note the longer time oscillations seen in Fig.~9 as compared to the results in Fig.~8.

\begin{figure}
\includegraphics[width=8.0cm,height=8.0cm]{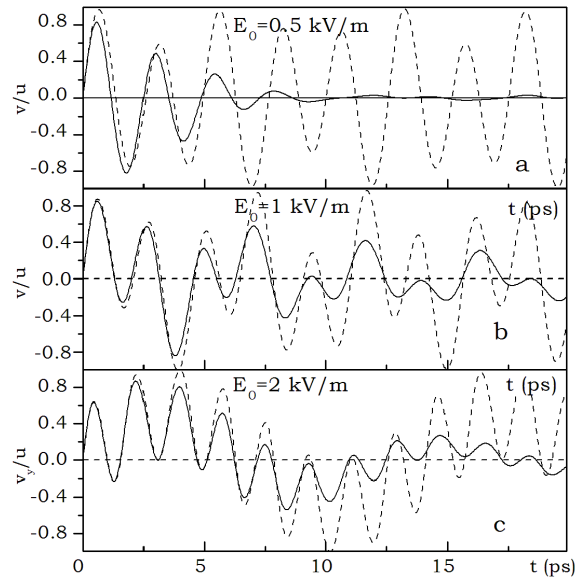}
\caption{Average velocity~$\langle v_x(t) \rangle$ of Gaussian packet (solid lines) compared with corresponding velocity of delta-like
         packet (dashed lines) having the same initial momentum~$\hbar k_{0y}$ for three field intensities.
         Driving frequency is~$\omega_D=2\times 10^{12}$~s$^{-1}$. Packet parameters are the same
         as in Fig.~\ref{Fig8}. It is seen that the finite width of the packet does not change
         oscillation frequency.} \label{Fig10}
\end{figure}

\begin{figure}
\includegraphics[width=8.0cm,height=8.0cm]{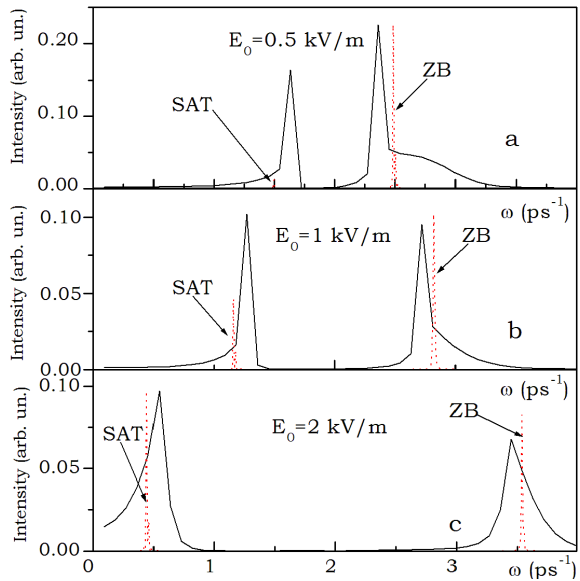}
\caption{Frequency spectrum (not normalized) for Gaussian wave packet (solid lines) and delta-like packet (dotted lines) for three
         field intensities. Driving frequency is~$\omega_D=2\times 10^{12}$~s$^{-1}$.
         Packet parameters are the same as in Fig.~\ref{Fig8}. For~$E_0= 0.5$ kV/m the intensity of satellite component
         is very small.} \label{Fig11}
\end{figure}

In Fig.~10 we compare the MZB oscillations for a Gaussian packet centered around~${\bm k}_0=(0,k_{0y})$ (solid lines)
with those for a delta-like packet having the same wave vector~${\bm k}_0$ (dashed lines) for three values of electric field.
The important feature of the decaying patterns is their nearly unchanged frequency, clearly visible in Fig.~10.
The disappearance of the packet motion in time is characteristic of the wave packets~\cite{Lock1979}.
It is worth noting that the decay times of ZB-like oscillations shown in
Figs.~8,~9 and~10 are in the scale of picoseconds, whereas those shown in Ref.~\cite{Rusin2007b} were in the scale of femtoseconds.
The reason for this difference is that the widths of wave packets used in Ref.~\cite{Rusin2007b} were
on the order of~$d\simeq~40$ \AA, while the widths used above are on the
order~$d\simeq~4$~$\mu$m. The general rule is that the wider the packet (i.e. narrower in the~$k$-space),
the longer the decay time of ZB oscillations.

In order to find precisely the frequency components of the motion we calculate the Fourier transform of the average packet
velocity
\begin{equation}
 \langle v_x(\omega)\rangle = \int_{-\infty}^{\infty} \langle v_x(t)\rangle e^{i\omega t} dt,
\end{equation}
as well as the intensities of Fourier components~$I_x(\omega) \propto |\langle v_x(\omega)\rangle|^2$.
The results are shown in Fig.~11 by solid lines. They are compared to the power spectrum of a delta-like packet
having the same wave vectors~$k_{0y}=1.2 \mu$m$^{-1}$ and~$k_{0x}=0$ and marked with the delta-like dotted peaks.
It is seen that the frequencies for the delta-like packets and the maxima for the Gaussian packets
are close to each other for all electric fields.

\subsection{Sub-packets. Trajectories}

It was shown previously that, in the field-free case, the fast decay of ZB oscillations results from a separation
of sub-packets containing positive and negative energy states~\cite{Rusin2007b}.
Here we carry a corresponding analysis for the time-dependent Hamiltonian in Eq.~(\ref{GH}).
For every instant of time this Hamiltonian can be
expanded in the basis of its eigenstates~$|1\rangle$ and~$|2\rangle$
\begin{equation} \label{Sub_H}
 \hH(t)= \lambda|1\rangle\langle 1| - \lambda|2\rangle\langle 2|,
\end{equation}
where~$\lambda=+\sqrt{H_{21}H_{12}}$ and
\begin{equation} \label{Sub_12}
|1\rangle = \frac{1}{\sqrt{2}}\left( \begin{array}{c} 1 \\ H_{21}/\lambda \end{array}\right), \hspace*{1em}
|2\rangle = \frac{1}{\sqrt{2}}\left( \begin{array}{c} -H_{12}/\lambda \\ 1 \end{array}\right).
\end{equation}
In the above expressions we treat~$\hat{\bm p}$ as a c-number:~${\bm p}=\hbar {\bm k}$. Since the elements~$H_{21}$ and~$H_{12}$ of
the Hamiltonian in Eq.~(\ref{GH}) depend on time, also~$\lambda$,~$|1\rangle$ and~$|2\rangle$
are functions of time. For~$E_0 \rightarrow 0$ the eigenvalue~$\lambda(t)$ reduces to the electron energy in
graphene:~$\epsilon_k=u\hbar|k|$. In the absence of fields the states~$|1\rangle$ and~$|2\rangle$
reduce to the conduction and valence bands of graphene, respectively. At any instant of time one can expand
the wave packet~$\Psi(t)$
\begin{equation} \label{Sub_Psi}
 \Psi(t) = a_1|1\rangle + a_2|2\rangle,
\end{equation}
where~$a_1=\langle 1|\Psi\rangle$ and~$a_2=\langle 2|\Psi\rangle$. Then we can expand the average packet velocity using the
states~$|1\rangle$ and~$|2\rangle$ and obtain from Eq.~(\ref{Gvx})
\begin{eqnarray} \label{Sub_v12}
\langle v_x(t) \rangle &=& |a_1|^2\langle 1|u\sigma_x| 1\rangle + |a_2|^2\langle 2|u\sigma_x| 2\rangle \nonumber\\
                       &+& a_1^*a_2\langle 1|u\sigma_x| 2 \rangle + a_2^*a_1\langle 2|u\sigma_x| 1 \rangle \nonumber \\
                       &\equiv& \langle v_x^{11}(t) \rangle + \langle v_x^{22}(t) \rangle + \langle v_x^{12}(t) \rangle+ \langle v_x^{21}(t) \rangle. \ \
\end{eqnarray}

\begin{figure}
\includegraphics[width=8.0cm,height=8.0cm]{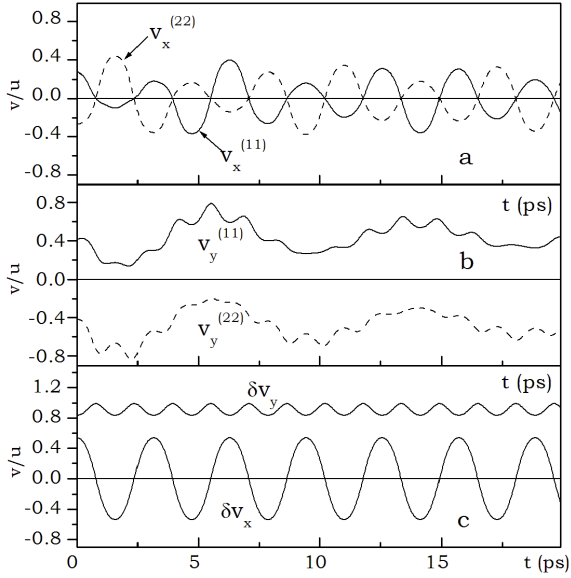}
\caption{Average velocities of sub-packets corresponding to positive and negative energy states; a) motion of sub-packets in the~$x$
         direction; b) motion of sub-packets in the~$y$ direction; c) relative velocity between the sub-packets.
         Electric field is~$E_0=1$ kV/m and~$\omega_D=2\times 10^{12}$~s$^{-1}$.
         Packet parameters are the same as in Fig.~\ref{Fig8}.} \label{Fig12}
\end{figure}

Terms in the first line of Eq.~(\ref{Sub_v12}) describe the motion of centers of the two sub-packets,
first with the positive and second with the negative eigenvalue~$\lambda$. In the field-free case these terms lead to a rectilinear motion
of sub-packets centers. In the presence of a driving wave these terms oscillate. Two terms in the
second line of Eq.~(\ref{Sub_v12})
describe an interference between the sub-packets. In the field-free case these terms are responsible for the ZB oscillations,
and the same occurs in the presence of a driving field. These terms are nonzero when the sub-packets are close together,
when they move away from each other the oscillations disappear. In Fig.~12a we plot the motion of sub-packet centers
for~$\langle v_x(t) \rangle$ calculated in Eq.~(\ref{Sub_v12}). The velocities of sub-packets oscillate
around~$\langle v_x \rangle=0$ with similar frequencies but opposite phases.
Thus, in the~$x$ direction the sub-packets do not move away from each other.

In Fig.~12b we plot velocities of the two sub-packets in the~$y$ direction using formulas analogous to that in Eq.~(\ref{Sub_v12}).
The motion is different from that in the~$x$ direction. The velocities of sub-packets have different signs
which means that they move in opposite directions. In Fig.~12c we show relative velocities of
both sub-packets in~$x$ and~$y$ directions. The relative velocity in the~$x$ direction oscillates around zero,
which means that the sub-packets are close to each other,
while the relative velocity in the~$y$ direction is nearly constant with small superimposed oscillations.
Thus the sub-packets move away each from other, cease to overlap and the MZB oscillations disappear similarly
to the field-free case~\cite{Rusin2007b}.

\begin{figure}
\includegraphics[width=8.0cm,height=8.0cm]{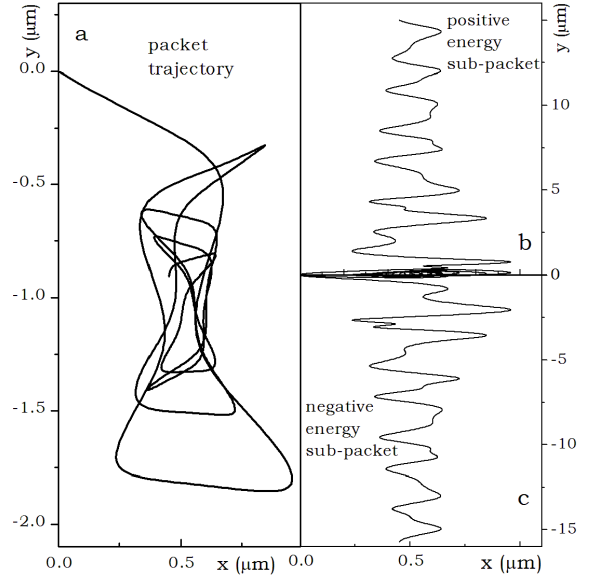}
\caption{Trajectories of the packet and sub-packets consisting of states with positive and negative energies during first 35 ps
         of motion. a) trajectory of the packet;
         b) and c) trajectories of sub-packet having states positive (negative) energies.
         Electric field is~$E_0=1$ kV/m and~$\omega_D=2\times 10^{12}$~s$^{-1}$.
         Packet parameters are the same as in Fig.~\ref{Fig8}.} \label{Fig13}
\end{figure}

Packet trajectory can be calculated in two alternative ways. We can calculate the trajectory
as an integral over the average velocity
\begin{equation} \label{Sub_r}
 \langle {\bm r}(t) \rangle = \int_0^t \langle {\bm v}(t') \rangle dt'.
\end{equation}
Alternatively, one can calculate the trajectory directly
\begin{equation} \label{Sub_rPsi}
\langle {\bm r}(t) \rangle = \langle\Psi(t)|{\bm r}|\Psi(t)\rangle.
\end{equation}
To find the trajectory in the~$x$ direction we insert twice the unity operator and obtain
\begin{eqnarray} \label{Sub_xPsi}
 \langle x(t) \rangle &=& \sum_{{\bm k}{\bm k}'}\langle \Psi|{\bm k}'\rangle\langle {\bm k}'|x|{\bm k}\rangle\langle {\bm k}|\Psi\rangle \nonumber \\
        &=& \frac{1}{4\pi^2} \int d^2 {\bm r} \int d^2{\bm k}'d^2{\bm k} \Psi_{\bm k'}^{\dagger}\Psi_{\bm k}
    \left( e^{i{\bm r}{\bm k}} x e^{-i{\bm r}{\bm k'}}\right), \ \ \
\end{eqnarray}
where~$\Psi_{\bm k}$ is the solution of Eq.~(\ref{GHw}) for given~${\bm k}=(k_x,k_y)$ and the dagger means the Hermitian conjugate.
If the function~$\Psi_{\bm k}$ tends sufficiently fast to zero for large~${\bm k}$ (e.g. exponentially),
one can change the order of integration in Eq.~(\ref{Sub_xPsi}) and obtain
\begin{eqnarray} \label{Sub_x}
 \langle x(t) \rangle &=& \frac{i}{4\pi^2} \int \Psi_{\bm k}^{\dagger}(t) \frac{\partial \Psi_{\bm k}(t)}{\partial k_x} d^2{\bm k}, \\
                \label{Sub_y}
 \langle y(t) \rangle &=& \frac{i}{4\pi^2} \int \Psi_{\bm k}^{\dagger}(t) \frac{\partial \Psi_{\bm k}(t)}{\partial k_y} d^2{\bm k}.
\end{eqnarray}
Equations~(\ref{Sub_x}) and~(\ref{Sub_y}) give~$\langle x(t) \rangle$ and~$\langle y(t) \rangle$ at any instant of time,
while Eq.~(\ref{Sub_r}) requires knowledge of~$\langle {\bm v}(t')\rangle$ for all~$t'\in (0,t)$. Choice of one of the two methods
depends on details of the numerical procedure. For delta-like packets, Eq.~(\ref{Sub_r}) is more convenient because
one does not deal with differentiation of the Dirac delta functions in Eqs.~(\ref{Sub_x}) and~(\ref{Sub_y}).

In Fig.~13a we show the calculated packet trajectory for the first 35 ps of motion and
in Figs. 13b and 13c the trajectories of the two sub-packets.
It is seen that the packet trajectory begins at~${\bm r}(0)={\bm 0}$
and the packet moves in an irregular slowly converging orbit. In the field-free case the packet center is displaced
only in the~$x$ direction. The motion of sub-packets, as indicated in Figs.~13b and~13c, is different.
The sub-packets move in opposite directions, so that their centers move away each from other. After 35~ps each
of the sub-packets is displaced about 15~$\mu$m i.e.
the distance between their centers exceeds the packet width~$d$=4~$\mu$m. The results presented in Fig.~13 indicate that the mechanism
responsible for the disappearance of MZB is similar to the field-free case~\cite{Rusin2007b}.

\section{Medium polarization}

\begin{figure}
\includegraphics[width=8.0cm,height=8.0cm]{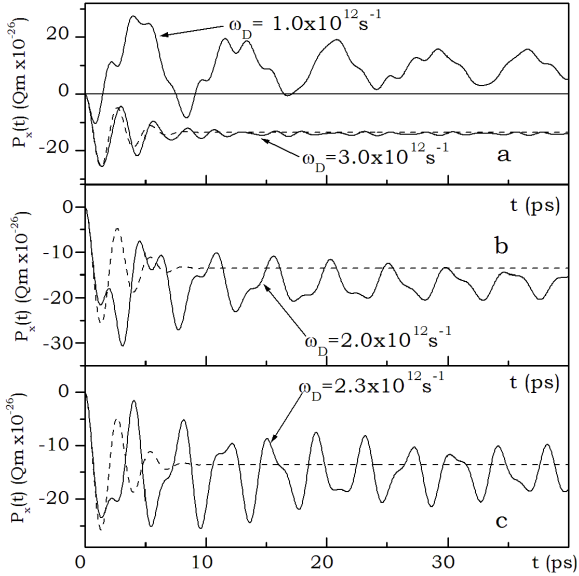}
\caption{Polarization~$P_x$ vs. time for four driving frequencies~$\omega_D$.
         Dashed line:~$P_x(t)$ in field-free case. Electric field is~$E_0=1$ kV/m.
         Packet parameters are the same as in Fig.~\ref{Fig8}.} \label{Fig14}
\end{figure}

Now we turn to the medium polarization caused by the driving wave.
The time-dependent polarization~${\bm P}(t)$ induced by the wave packet is a physical quantity
measured frequently in the ultra-fast spectroscopy~\cite{MukamelBook}.
Various experimental techniques allow one to determine packet motion from polarization oscillations
in molecules~\cite{ZewailNobel1999}, and quantum wells or superlattices~\cite{Martini1996,Lyssenko1996}. The polarization is defined as
\begin{equation} \label{EMQ_Px}
 P_x(t) = -e\langle \Psi(t)|x|\Psi(t)\rangle,
\end{equation}
and similarly for~$P_y(t)$. Thus~${\bm P}(t)$ is proportional to the average packet position.
The explicit forms of~${\bm P}(t)$
can be obtained from Eq.~(\ref{Sub_r}) or, equivalently, from Eqs.~(\ref{Sub_x}) and~(\ref{Sub_y}).
In the time-dependent spectroscopy one expands polarization in the power series in field intensity~${\bm E}(t)$ \cite{BoydBook}
\begin{eqnarray} \label{EMQ_P123}
 {\bm P}(t) &=& \epsilon_0\{\chi^{(1)}{\bm E}(t) + \chi^{(2)}{\bm E}^2(t) + \chi^{(3)}{\bm E}^3(t) + \ldots \} \nonumber \\
            &=&{\bm P}^{(1)}(t) + {\bm P}^{(2)}(t)+{\bm P}^{(3)}(t) + \ldots,
\end{eqnarray}
where~$\chi^{(1)}$ is the linear susceptibility and~$\chi^{(2)}$ and~$\chi^{(3)}$ are non-linear susceptibilities of
the second and third order, respectively. The above expansion assumes that the polarization depends on the instantaneous
value of the electric field which implies that the system is lossless and dispersion-less~\cite{BoydBook}.
In experiments, one can measure the linear polarization as well as the higher-order nonlinear polarizations using
various techniques, e.g. photon echo or pump-and-probe measurements, see Ref.~\cite{MukamelBook}.
In our approach we obtain in Eq.~(\ref{EMQ_P123}) an exact form of~${\bm P}(t)$ including {\it all}
expansion orders. Connection between the third order polarizations observed in photon-echo experiments
and the complete polarization, as given in Eq.~(\ref{EMQ_P123}), is discussed in Refs.~\cite{Ebel1994,Seidner1995}.

In Fig.~14 we plot the time-dependent polarization~$P_x(t)$ calculated for one electron prepared in the form
of a Gaussian packet for five driving frequencies~$\omega_D$. The derivatives of~$\Psi(t)$ with
respect to~$k_x$ and~$k_y$ are calculated numerically using the four-point differentiation rule.
Dashed lines indicate polarization for~$E_0=0$. In Fig.~14 we see different regimes of MZB oscillations:
one frequency for large-driving frequency, two frequencies in the~$\omega_D\simeq\omega_Z$ regime and
many frequencies for low~$\omega_D$.
In Fig.~14a one can see that the polarization for large driving frequency nearly equals that for~$\omega_D=0$.
Thus for large~$\omega_D$ the rapid oscillations of the driving wave average out to zero and they do not alter the electron motion.
For~$\omega_D$ close or smaller than~$\omega_Z$ the decay time of polarization oscillations is
about ten times longer than in the field-free case, which makes possible experimental observations easier.
The decay time of~$P_x(t)$ oscillations is longest for~$\omega_D\simeq\omega_Z$, see Fig.~14c.

\begin{figure}
\includegraphics[width=8.0cm,height=8.0cm]{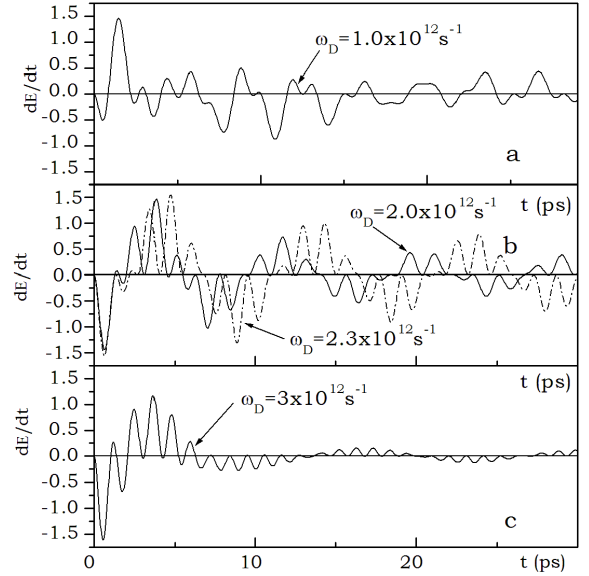}
\caption{Instantaneous power~$d{\cal E}/dt$ vs. time for three driving frequencies~$\omega_D$.
         Electric field is~$E_0=1$ kV/m.
         Packet parameters are the same as in Figure~\ref{Fig8}.} \label{Fig15}
\end{figure}

In the present work we concentrate on the average packet velocity. Now we analyze the
observable quantity which allows one, at least in principle, to measure packet velocity in a direct way.
Let us consider the time-dependent energy of the system:~${\cal E}=\langle \Psi|\hH|\Psi\rangle$~\cite{Ebel1994,Seidner1995}.
The change of~${\cal E}$ in time, i.e. the instantaneous power emitted or absorbed by the system is
\begin{eqnarray} \label{EMQ_I}
 I= \frac{d{\cal E} }{dt} &=& \langle \frac{d\Psi}{dt}|\hH|\Psi\rangle + \langle \Psi|\hH|\frac{d\Psi}{dt} \rangle +
   \langle \Psi|\frac{\partial \hH}{\partial t}|\Psi\rangle \nonumber \\
  &=& \langle \Psi|\frac{\partial \hH}{\partial t}|\Psi\rangle = -|e| E(t) \langle v_x(t)\rangle.
\end{eqnarray}
It is seen that the instantaneous power is proportional to the product of the electric field and the
average velocity~$\langle v_x(t)\rangle$.
In Figure~15 we plot this power for four values of electric field. The results resemble those for~$\langle P_x(t)\rangle$
but more frequencies appear. Note that, in experiments, it is more convenient
to measure time-dependent polarization rather than the instantaneous power emitted or absorbed by the system.
Thus the theoretical average velocity must be integrated over time in order to obtain the average position,
see Eqs.~(\ref{Sub_r}) and~(\ref{EMQ_Px}).

\section{Discussion}

As we mention in the Introduction, our idea of considering the
Multi-mode Zitterbewegung, i.e. the trembling motion of electrons in a
periodic potential in an additional presence of a driving
electromagnetic wave, was inspired by considerations of
Super-Bloch-Oscillator~\cite{Thommen2002,Kolovsky2009,Haller2010}.
The two systems seem analogous, since both
are characterized by an internal electron frequency related to the
periodic lattice potential and in both the electrons are subjected in
addition to the interaction with a driving wave. However, it turned out
that our results do not much resemble those of the
Super-Bloch-Oscillator. Rather, they are related to descriptions of
systems consisting of two levels in the presence of a light wave i.e.
to the problems in quantum optics. This affinity is expressed by similar
mathematical elements, like the Heun function~\cite{Heun2010} and the Mathieu
equation mentioned above.

As to the quantum optics, it usually deals
with levels, whereas we deal with bands, and it is mostly concerned with
the population of levels, while we are interested in the electron
motion. In quantum optics, one of the problems is that of gauge i.e.
should the radiation be introduced by the electric scalar potential or
by the vector potential. In our approach, we introduce the wave using
the vector potential in the electric dipole approximation. This is a
convenient choice and it leads to no ambiguities since we obtain exact
numerical solutions which should be independent of the gauge.

We restrict our considerations to wave intensities~$E_0 < 1\times 10^6$~V/m
in order to avoid considerations of electron emission by the field. In
this connection on should mention two issues. The first are the initial
conditions for our treatment. We do not consider interband excitations of
electrons in graphene by the incoming wave but assume from the beginning
that the electron wave function has higher and lower components,
see Eq.~(\ref{GPsi0}). This tacitly assumes that the Fermi level is located in the valence
band of graphene, so that the electron wave packet can contain both
components and their interference results in the Zitterbewegung.

The second issue is concerned with a possible emission of radiation. During
the ``classical'' trembling motion in a solid the electron does not radiate
because it is in the Bloch eigenenergy state. However, once the electron
is additionally driven by an external wave, it is clearly not in the
eigenenergy state and it can radiate. We do not consider here this
radiation, we only consider a polarization of the graphene medium caused
by the resulting motion. Also, our theory is a one-electron approach and
we do not consider effects related to the fact that various electrons may
oscillate with different phases.

We also neglect the effect of finite temperature on the electron oscillations.
The driving frequency~$\omega_D=2\times 10^{12}$~s$^{-1}$, which is often used
in our paper, corresponds to the temperature~$T_D=\hbar \omega_D/k_B \simeq 15.3$~K, so the experiments
measuring the MZB should be performed in the liquid helium temperatures.
A finite temperature would also lead, via the electron-phonon interaction, to a faster
decay of MZB oscillations.

Another limiting factor of our approach is concerned with initial values
of the wave vector characterizing the wave packet. This problem is present
in {\it almost all} recent papers on Zitterbewegung, beginning with that of
Schliemann {\it et al.}~\cite{Schliemann05}.
When one considers an electron localized in the form od a wave packet and
takes its initial components~$(1, 0)^T$, see Eq.~(\ref{GPsi0}),
it turns out that in order to have ZB in one direction one needs
to have a non-vanishing initial momentum in the perpendicular direction.
The question arises, how to create such momentum. If, which is tacitly
assumed, the packet is created by a flash of laser light, the electron
will have a very small initial momentum since that carried by light is
small. An alternative way to create a sizable electron momentum would be
to excite the electron by an acoustic phonon, but then the electron energy
will be rather small. Still, in the simulation of the 1+1 Dirac equation
by cold ions interacting with laser beams with the resulting ZB, as
realized by Gerritsma {\it et al.}~\cite{Gerritsma2010}, a non-vanishing momentum of the wave
packet~$(1,1)^T$ was created in the same direction.
This suggests that the conditions considered in our
approach could be created by an appropriate simulation. A system, in which
a non-negligible initial momentum is present, are carbon nanotubes since
there always exists a built-in quantized momentum in the direction
perpendicular to the tube's axis~\cite{Rusin2007b,Zawadzki2006}.
It appears that, instead of taking {\it a priori} a packet in a given form,
it would be more realistic to determine what packet forms are created in specific experiments and use
these forms in subsequent calculations.

It is worth emphasizing that the two applied approximations, namely the
rotating-wave and high driving frequency approximations give good results
in two different regimes. The RWA applies to resonance conditions~$\omega_D\simeq \omega_Z$,
while HDF applies to~$\omega_D\gg \omega_Z$ regime.
As far as RWA is concerned the adopted procedure neglects frequencies
away from the resonance, while in the HDF regime the condition
of~$\kappa=\omega_e/\omega_D \ll 1/2$ is well satisfied,
see Eqs.~(\ref{HDF_Schb1}),~(\ref{HDF_Schb1}) and~(\ref{HDF_kappa}).

Finally, it should be mentioned that San Roman {\it et al.}~\cite{Roman2003} described
the Zitterbewegung of Dirac electrons driven by an intense laser field.
According to a close analogy between relativistic Dirac electrons in a
vacuum and electrons in narrow-semiconductors, see
Ref.~\cite{ZawadzkiHMF}, this
problem resembles ours since electrons in graphene, having a linear
relation between the energy and momentum, represent the so called extreme
relativistic case. However, the treatments in the two cases are different.
The reason is that for the Dirac equation there exist Volkov solutions.
They exist because the light velocity~$c$ is the same as the maximum
velocity of electrons. This is not the case for electrons in solids: the
maximum velocity for electrons in graphene is around~300 times smaller
than the light velocity. The description of Ref.~\cite{Roman2003} for the
Dirac electrons shows that the main
effect of laser light is caused by its magnetic component, resulting in a
collapse-and-revival pattern of ZB oscillations. This resembles the
situation in graphene,
as it has been shown that ZB in the presence of an external
magnetic field exhibits the-collapse-and revival pattern, see
Refs.~\cite{Rusin2008,Romera2009}.
On the other hand, since in the present paper we have neglected the
magnetic component of the electromagnetic wave [see remarks after
Eq.~(\ref{GH})] the effect of collapse-and-revival does not appear and we
concentrate on the multi-mode behavior.

\section{Summary}
Electrons in monolayer graphene in the presence of an electromagnetic (or
electric) wave are considered. It is shown that, in addition to the
Zitterbewegung oscillations related to the periodic potential of
graphene lattice, the electron interaction with the driving external wave
gives rise to hybrid oscillations involving both the ZB frequency~$\omega_Z$
and driving frequency~$\omega_D$. Three distinct regimes of the motion are
identified: one-frequency mode for~$\omega_D \gg \omega_Z$, two-frequency mode
for~$\omega_D \simeq \omega_Z$, multi-frequency mode for~$\omega_D \ll \omega_Z$,
resulting from inherent nonlinearity of the system. Also, the presence of
driving wave activates additional oscillation directions, not present in
pure ZB motion. Dependence of the mode behavior on the intensity of
driving wave is investigated. Electrons are described by either delta-like
or Gaussian wave packets and it is shown that the essential oscillation
characteristics (frequency and amplitude) depend only weakly on the packet
width. It is indicated that the presence of a driving wave should
facilitate observations of electron Zitterbewegung in semiconductors for
two reasons: 1) It gives an additional external parameter for varying
ZB-related frequency, 2) it prolongs the decay time of hybrid
oscillations, as compared to that of pure ZB oscillations.

\appendix

\section{Classical description}

Here we briefly consider a classical description of a
one-dimensional electron motion in the presence of a periodic potential
and a driving electric wave. Let the periodic
potential of the lattice be given by~$W(x)=W_0\sin(2\pi x/a)$, where~$a$ is the lattice
period, and the interaction with the wave described by~$eE_0x\cos(\omega_Dt)$.
Then the second Newton law of motion reads
\begin{equation} \label{App_Class1}
 m_0\frac{dv}{dt}=-W_0\left(\frac{2\pi}{a}\right) \cos\left(\frac{2\pi x}{a}\right) + eE_0\cos(\omega_Dt),
\end{equation}
where~$m_0$ is the electron mass. We find approximate solutions of the above
equation by iteration. In the initial step the
velocity is taken to be constant and equal to~$v_0$. This gives~$x_0=v_0t$. By
putting~$x_0$ into Eq.~(\ref{App_Class1}) and integrating over time one obtains the first
approximation to velocity
\begin{equation} \label{App_Class2}
v_1(t)= -A_1\sin(\omega_Zt) + B_1\sin(\omega_Dt),
\end{equation}
where~$\omega_Z=2\pi v_0/a$ is the frequency of Zitterbewegung [we put the initial
value~$v_1(0)=0$], and~$A_1$,~$B_1$ are constants.
The first term in Eq.~(\ref{App_Class2}) describes oscillatory deviations
of velocity from its average value~$v_0$ due to ZB and the second accounts
for the corresponding contribution of the wave. In this approximation there
is no mixing of frequencies. Now one can again integrate the velocity of Eq.~(\ref{App_Class2})
over time to obtain~$x_1(t)$ and put it back to the initial
Eq.~(\ref{App_Class1}). This gives
\begin{eqnarray} \label{App_Class3}
m_0 \frac{dv_2}{dt} &=& -W_0\frac{2\pi}{a}\cos\left[A_1\cos(\omega_Zt)+B_1\cos(\omega_Dt) \right] \nonumber \\
                  &&+ eE_0\cos(\omega_Dt).
\end{eqnarray}
It is seen that in the second approximation there is mixing of
frequencies~$\omega_Z$ and~$\omega_D$. This mixing is a general feature of our
nonlinear problem, it is seen in numerous expressions of the quantum
treatment, see e.g. Eq.~(\ref{RWA_vx_w}). In Eq.~(\ref{HDF_vy}) one can also see a
characteristic trigonometric function of a trigonometric function,
appearing in the above classical expression~(\ref{App_Class3}).

\section{MZB vs. quantum optics}

We find a correspondence between average velocities~$\langle v_x(t) \rangle$ and~$\langle v_y(t) \rangle$
in graphene and the probability of occupation of the upper
level in the two-level model driven by a periodic force. This problem is frequently considered in quantum optics.
The two-level model is usually described by the Schrodinger equation
\begin{equation} \label{A2L_H2Lw}
 i\frac{\partial \Phi}{\partial t} = \left\{\left[f_0 + f_1\cos(\omega_Dt)\right]\sigma_x + \frac{\nu}{2}\sigma_z \right\}\Phi.
\end{equation}
It has been observed that solutions of the above equation can be expressed in terms of Heun functions~\cite{Heun2010}.
By introducing the unitary operator~$\hS_x=(1-i\sigma_x)/\sqrt{2}$,
acting with~$\hS_x$ on both sides of Eq.~(\ref{GHw}) and using~$1=\hS_x^{-1}\hS_x$, one obtains
\begin{equation} \label{A2L_HwS}
 i\frac{\partial \Phi}{\partial t} = \left\{[uk_x + \omega_e \cos(\omega_Dt)]\sigma_x + uk_y\sigma_z \right\}\Phi,
\end{equation}
where~$\Phi=\hS_x\Psi$. Setting~$f_0=uk_x$,~$\nu=2uk_y$, and~$f_1=\omega_e$ we transform
the Hamiltonian of Eq.~(\ref{GHw}) into the Hamiltonian of the two-level model~(\ref{A2L_H2Lw}).

Let~$\Phi(t)=(\Phi_1(t),\Phi_2(t))^T$.
Then the probability of occupation of the upper level is~$P^{up}=|\Phi_1(t)|^2$.
On the other hand, using the projection operator~$\hP$ of Eq.~(\ref{OA_hP})
the occupation probability of the upper state is~$P^{up}=\langle \Phi|\hP| \Phi \rangle$.
By using~$\Phi=\hS_x\Psi$ we find
\begin{equation} \label{A2L_Pup}
P^{up} = \frac{1}{2} + \frac{1}{2} \langle \Psi|\sigma_y| \Psi\rangle,
\end{equation}
which gives, see Eq.~(\ref{Gvx})
\begin{equation} \label{A2L_vx_Pup}
\langle v_y(t) \rangle = u \left(2P^{up}-1\right).
\end{equation}
The above equation together with Eq.~(\ref{A2L_HwS}) connects the results for~$P^{up}$ in the two-level model with the
Zitterbewegung in graphene in the~$y$ direction. Note that in quantum optics one calculates~$P^{up}$
taking~$\Phi(0)=(1,0)^T$, which gives~$\Psi(0)=(1,i)^T$, while in ZB calculations one takes~$\Psi(0)=(1,0)^T$.

\section{Relation to Rabi oscillations}

\begin{figure}
\includegraphics[width=8.0cm,height=8.0cm]{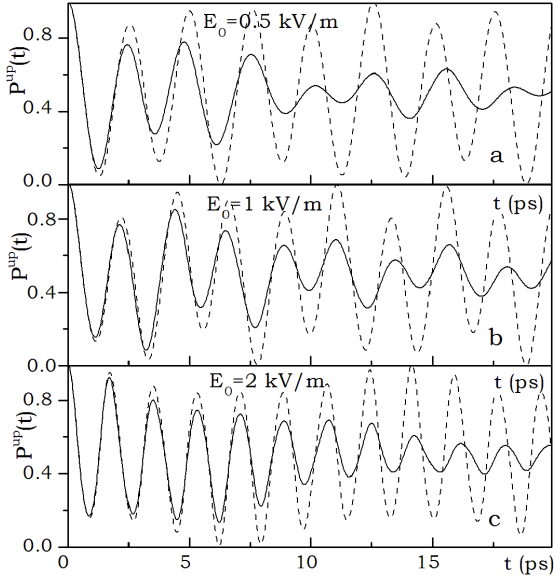}
\caption{Occupation probability of the upper energy branch for Gaussian packets (solid lines) and delta-like packets (dashed lines)
         for three electric fields and~$\omega_D=2\times 10^{12}$~s$^{-1}$. Dashed lines correspond to the Rabi oscillations in a two-level system.
         Packet parameters are the same as in Fig.~\ref{Fig8}.} \label{Fig16}
\end{figure}

Here we indicate relation between MZB and the Rabi oscillations, i.e. oscillations of the probability of
occupation of upper (lower) energy states. For the upper energy level the projection operator is
\begin{equation} \label{OA_hP}
\hP = \frac{1}{2}(1+\sigma_z),
\end{equation}
and the occupation probability of the upper state (conduction band) is
\begin{equation}
\langle \hP \rangle = \langle \Psi(t)|\hP| \Psi(t) \rangle,
\end{equation}
where~$\Psi(t)$ is the solution of the Schrodinger equation~(\ref{GHw}).
The probability~$\langle \hP \rangle$ can be calculated in the
way analogous to~$\langle v_x(t) \rangle$. For the rotating wave approximation we obtain
\begin{eqnarray} \label{RAB_P_RWA}
\langle \hP \rangle &=& \frac{1}{2} + \frac{1}{2} \left(1+\frac{\Delta}{\omega_R}\right)\cos(\omega_Dt+\omega_Rt ) \nonumber \\
                    && \ \ + \frac{1}{2} \left(1-\frac{\Delta}{\omega_R}\right)\cos(\omega_Dt-\omega_Rt ),
\end{eqnarray}
while for the large~$\omega_D$ approximation there is
\begin{eqnarray} \label{RAB_P_wd}
\langle \hP \rangle &=& \frac{1}{2} + \frac{1}{2}\cos(2\Lambda)\cos(\omega_Zt) \nonumber \\
                    &\simeq & \frac{1}{2} +\frac{1}{2}\cos(\omega_Zt).
\end{eqnarray}
Here~$\Delta=\omega_Z - \omega_D$ and~$\Lambda$ is given by Eq.~(\ref{HDF_phi_w}).
In Fig.~16 we show calculated Rabi oscillations for delta-like packets (dashed lines)
and Gaussian packets (solid lines) for
three values of external electric field. For the delta-like packets the oscillations are persistent while for
the Gaussian packets they decay in time. The decay is caused by sub-packets
moving in the opposite directions, as described above. The frequencies for delta-like packets are
nearly equal to those of Gaussian packets. We do not pursue this subject here since the Rabi oscillations
in graphene were treated in detail in Ref.~\cite{Vasko2010}.

\section{Perturbation and Fer-Magnus expansions}

\begin{figure}
\includegraphics[width=8.0cm,height=8.0cm]{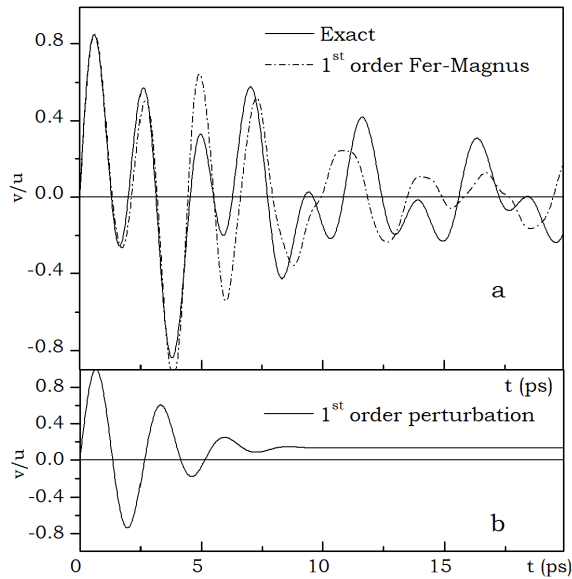}
\caption{Average velocity~$\langle v_x(t) \rangle$ of Gaussian packet calculated a) numerically (solid line),
         with the use of the Fer-Magnus expansion (dashed-dotted line) and b) using
         time-dependent perturbation.
         Electric field is~$E_0=1$ kV/m and~$\omega_D=2\times 10^{12}$~s$^{-1}$.
         Packet parameters are the same as in Fig.~\ref{Fig8}.} \label{Fig17}
\end{figure}

We compare our numerical time-dependent results with those obtained in the standard time-dependent perturbation
and in the Fer-Magnus expansions.
For small fields a natural way of finding approximate solutions of Eq.~(\ref{GHw}) is to treat the time-dependent term
as a perturbation.
Let~$\varphi_1$ and~$\varphi_2$ be the eigenstates of~$\hH_0$ corresponding to positive and negative energy states,
respectively, and~$\Psi_1(t), \Psi_2(t)$ be unknown eigenstates of complete~$\hH$.

We first express~$\Psi(t)$ as a linear combination of~$\Psi_1$ and~$\Psi_2$
\begin{equation} \label{AFM_Psi}
 \Psi(t) = a_1\Psi_1(t) + a_2\Psi_2(t),
\end{equation}
where~$a_1$ and~$a_2$ are time-independent coefficients to be found.
In the lowest order of time-dependent perturbation there is
\begin{equation} \label{AFM_Psi1}
 \Psi_1(t) \simeq \varphi_1e^{-i\omega_kt} + \frac{1}{i\hbar} \sum_{j=1,2} c_{1j}\varphi_je^{-is_j\omega_kt} + \ldots,
\end{equation}
where~$\omega_k=u\sqrt{k_x^2+k_y^2}$,~$s_1=+1$,~$s_2=-1$, and
\begin{equation} \label{AFM_c1j}
 c_{1j} = \langle \varphi_1|\omega_e\sigma_x|\varphi_j\rangle \int_0^t \cos(\omega_D t')e^{i\omega_k(s_1-s_j)t'} dt'.
\end{equation}
The expansion for~$\Psi_2(t)$ is analogous.
The coefficients~$a_j$ in Eq.~(\ref{AFM_Psi}) can be found from the initial condition~$\Psi(0)=(1,0)^T f({\bm k})$ for~$E_0=0$.
The matrix elements and the integration in Eq.~(\ref{AFM_c1j}) can be calculated analytically. Having the approximate form of~$\Psi(t)$
we can calculate the average velocity~$\langle v_x(t)\rangle$ for the wave packet. The results are shown in Fig.~17b. In Fig.~17a
we plot the corresponding exact results for~$\langle v_x(t)\rangle$ computed numerically.

Comparing the approximate results with the exact ones it can be seen that, in this problem,
the first order time-dependent perturbation presents a poor approximation.
First, for long times the perturbation gives a constant, nonzero velocity,
while the exact results give vanishing velocity.
Second, the decay time of MZB calculated by the perturbation is too short. Still, the perturbation expansion gives
proper oscillation frequency and correct behavior of~$\langle v_x(t)\rangle$ for short times.

Another possibility of obtaining approximate results is the Magnus expansion for the propagator of the Schrodinger equation
with time-dependent Hamiltonians~\cite{Magnus1954}. Having exact or approximate propagator~$\hU(t)$ one can calculate~$\Psi(t)=\hU(t)\Psi(0)$
and then the average velocity, see Eqs.~(\ref{Gvx}) and~(\ref{Gvy}).
In our approach we examine the Fer-Magnus~(FM) expansion, which is more suitable for problems
periodic in time~\cite{Fer1958}. The advantage of Magnus or FM expansions is
that the propagator in Eq.~(\ref{AFM_FM}) remains unitary in {\it every} order of perturbation, which ensures the proper norm of the wave function.

Following the suggestion of Blanes {\it et al.}~\cite{Blanes2009} we transform Eq.~(\ref{GHw}) to
the interaction picture~$ i\hbar d\Psi^i/dt = V^i(t)\Psi^i$,
where~$\Psi^i= e^{+i\hH_0 t/\hbar}\Psi$ and~$V^i(t)=e^{i\hH_0t/\hbar} V(t)e^{-i\hH_0t/\hbar}$.
In the Fer-Magnus expansion the propagator is assumed in the form
\begin{equation} \label{AFM_FM}
 \hU = \exp\left(-i\sum_{n=1}^{\infty} \hat{\Lambda}_n(t)\right) \exp\left(-it\sum_{n=1}^{\infty} \hat{F}_n\right),
\end{equation}
where~$\hat{\Lambda}_n(t)$ and~$\hat{F}_n$ are given by recursive formulas~\cite{Blanes2009}.
At present there exist no closed formulas for~$\hat{F}_n$ and~$\hat{\Lambda_n}(t)$ with arbitrary~$n$.
For~$n=1$ there is
\begin{eqnarray} \label{AFM_FM1}
 \hat{F}_1          &=& \frac{1}{T}\int_0^T V^{int}(t') dt', \\
 \hat{\Lambda_1}(t) &=& \int V^{int}(t') dt' - t\hat{F_1}.
\end{eqnarray}
Expressions for~$n=2,3$ are given in Ref.~\cite{Mananga2011}.
Having an approximate propagator~$\hU(t)$ for~$n=1$ we calculate average packet velocity and the results
are plotted in Fig.~17a by dash-dotted line.
The FM method correctly approximates frequencies and amplitudes of MZB motion both for short and long times.
For short times the agreement is very good, for longer times a shift of phase appears.
The decay time calculated with the use of FM agrees quite well with that obtained numerically and
the vanishing velocity is in agreement with the exact results.
We conclude that the FM expansion can be applied to calculations of the MZB motion.


\begin{thebibliography}{99}

\bibitem{Schroedinger1930} E. Schrodinger, Sitzungsber. Preuss. Akad.
                          Wiss. Phys. Math. Kl. {\bf 24}, 418 (1930).
                          Schrodinger's derivation is reproduced in
                          A. O. Barut and A. J. Bracken, Phys. Rev. D {\bf 23}, 2454 (1981).
\bibitem{Gerritsma2010}   R. Gerritsma, G. Kirchmair, F. Zahringer, E. Solano, R. Blatt, and C. F. Roos,
                          Nature {\bf 463}, 68 (2010).
\bibitem{Zawadzki2011}    W. Zawadzki and T. M. Rusin, J. Phys. Cond. Matt. {\bf 23}, 143201 (2011).
\bibitem{Zawadzki2005KP}  W. Zawadzki, Phys. Rev. B {\bf 72}, 085217 (2005).
\bibitem{Schliemann05}    J. Schliemann, D. Loss, and R. M. Westervelt, Phys. Rev. Lett. {\bf 94}, 206801 (2005).
\bibitem{Zawadzki2010}    W. Zawadzki and T. M. Rusin, Phys. Lett. A {\bf 374}, 3533 (2010).
\bibitem{Zawadzki2013}    W. Zawadzki, Acta Phys. Pol. 123, 132 (2013).

\bibitem{Katsnelson2006}  M. I. Katsnelson, Europ. Phys. J. B {\bf 51} 157, (2006).
\bibitem{Cserti2006}      J. Cserti and G. David, Phys. Rev. B {\bf 74} 172305, (2006).
\bibitem{Maksimova2008}   G. M. Maksimova, V. Ya. Demikhovskii, and E. V. Frolova, Phys. Rev. B {\bf 78}, 235321 (2008).
\bibitem{David2010}       G. David and J. Cserti, Phys. Rev. B {\bf 81} 121417(R), (2010).
\bibitem{Rusin2007b}      T. M. Rusin and W. Zawadzki, Phys. Rev. B {\bf 76}, 195439 (2007).
\bibitem{Wallace1947}     P. R. Wallace, Phys. Rev. {\bf 71}, 622 (1947).
\bibitem{Slonczewski1958} J. C. Slonczewski and P. R. Weiss, Phys. Rev. {\bf 109}, 272 (1958).
\bibitem{McClure1956}     J. W. McClure, Phys. Rev. {\bf 104}, 666 (1956).
\bibitem{Novoselov2005}   K. S. Novoselov, A. K. Geim, S. V. Morozov, D. Jiang, M. I. Katsnelson,
                          I. V. Grigorieva, S. V. Dubonos, and A. A. Firsov, Nature {\bf 438}, 197 (2005).

\bibitem{Thommen2002}     Q. Thommen, J. C. Garreau and V. Zehnle, Phys. Rev. A {\bf 65}, 053406 (2002).
\bibitem{Kolovsky2009}    A. Kolovsky and H.J. Korsch, arXiv:0912.2587 (2009).
\bibitem{Haller2010}      E. Haller, R. Hart, M. J. Mark, J. G. Danzl, L. Reichsollner, and H. C. Nagerl,
                          Phys. Rev. Lett. {\bf 104}, 200403 (2010).

\bibitem{AllenBook}       L. Allen, and J. H. Eberly, {\it Optical Resonance and Two-level Atoms} (Wiley, New York, 1975).
\bibitem{SchleichBook}    W. P. Schleich, {\it Quantum Optics in Phase Space} (Wiley, Berlin, 2001).

\bibitem{Garraway1995}    B. M. Garraway and K. A. Suominen, Rep. Prog. Phys. {\bf 58}, 365 (1995).
\bibitem{Rabi1937}        I. I. Rabi, Phys. Rev. {\bf 51}, 652 (1937).
\bibitem{BoydBook}        R. W. Boyd, {\it Nonlinear Optics} (Academic Press, San Diego, 2003).
\bibitem{Lock1979}        J. A. Lock, Am. J. Phys. {\bf 47}, 797 (1979).

\bibitem{MukamelBook}     S. Mukamel, {\it Principles of nonlinear molecular spectroscopy} (Oxford University Press, New York, 1995).
\bibitem{ZewailNobel1999} A. Zewail, {\it Nobel lecture}, (1999); available on
                          http://nobelprize.org/nobel\_prizes/chemistry/ laureates/1999/zewail-lecture.pdf.

\bibitem{Martini1996}     R. Martini, G. Klose, H. G. Roskos, H. Kurz, H. T. Grahn, and R. Hey, Phys. Rev. B {\bf 54}, R14325 (1996).
\bibitem{Lyssenko1996}    V. G. Lyssenko, G. Valusis, F. Loser, T. Hasche, K. Leo,
                          M. M. Dignam, and K. Kohler, Phys. Rev. Lett. {\bf 79}, 301 (1997).

\bibitem{Ebel1994}        G. Ebel and R. Schinke, J. Chem. Phys. {\bf 101}, 1865 (1994).
\bibitem{Seidner1995}     L. Seidner, G. Stock, and W. Domcke, J. Chem. Phys. {\bf 103}, 3998 (1995).
\bibitem{Heun2010}        Q. Xie and W. Hai, Phys. Rev. A {\bf 82}, 032117 (2010).
\bibitem{Zawadzki2006}    W. Zawadzki, Phys. Rev. {\bf B} 74, 205439 (2006).
\bibitem{Roman2003}       J. San Roman, L. Roso, and L. Plaja, J. Phys. B: At. Mol. Opt. Phys. {\bf 36}, 2253 (2003).
\bibitem{ZawadzkiHMF}     W. Zawadzki, in {\it High Magnetic Fields in the Physics of Semiconductors II},
                          edited by G. Landwehr and W. Ossau (World Scientific, Singapore, 1997), p. 755.
\bibitem{Rusin2008}       T. M. Rusin and W. Zawadzki, Phys. Rev. B {\bf 78}, 125419 (2008).
\bibitem{Romera2009}      E. Romera and F. de los Santos, Phys. Rev. B {\bf 80}, 165416 (2009).
\bibitem{Vasko2010}       P. N. Romanets and F. T. Vasko, Phys. Rev. B {\bf 81}, 241411R (2010).

\bibitem{Magnus1954}      W. Magnus, Comm. Pure Appl. Math. {\bf 7}, 649 (1954).
\bibitem{Fer1958}         F. Fer, Bull. Classe Sci. Acad. Roy. Bel. {\bf 44}, 818 (1958).
\bibitem{Blanes2009}      S. Blanes, F. Casas, J. A. Oteo, and J. Ros, Phys. Rep. {\bf 470}, 151 (2009).
\bibitem{Mananga2011}     E. S. Mananga and T. Charpentier, J. Chem. Phys. {\bf 135}, 044109 (2011).

\end{thebibliography}
\end{document}